\documentclass[journal]{IEEEtran}

\usepackage{array}

\usepackage{enumerate}
\usepackage{subfigure}
\usepackage{url}
\usepackage[utf8]{inputenc}
\usepackage{cite}
\usepackage{amsmath,amssymb,amsfonts}
\usepackage{algorithmic}
\usepackage{graphicx}
\usepackage{textcomp}
\usepackage{stackengine}
\usepackage[dvipsnames]{xcolor}
\DeclareMathOperator{\sinc}{sinc}
\DeclareMathOperator{\E}{\mathbb{E}}
\def\BibTeX{{\rm B\kern-.05em{\sc i\kern-.025em b}\kern-.08em
    T\kern-.1667em\lower.7ex\hbox{E}\kern-.125emX}}
\ifCLASSINFOpdf
\else
\fi
\begin{document}
%
\title{Low-probability of Intercept/Detect (LPI/LPD) Secure Communications Using Antenna Arrays Employing Rapid Sidelobe Time Modulation}
%
%
%
\author{Jiahao~Zhao,
			Shichen~Qiao,~\IEEEmembership{Graduate Student Member,~IEEE,}
        John~H.Booske,~\IEEEmembership{Fellow,~IEEE,}
        and~Nader~Behdad,~\IEEEmembership{Fellow,~IEEE}
\thanks{
This work has been submitted to the IEEE for possible publication. Copyright may be transferred without notice, after which this version may no longer be accessible.

This work was supported in part by the McFarland-Bascom Professorship funds provided by the University of Wisconsin-Madison. \textit{(Corresponding author: Jiahao Zhao.)}%

J. Zhao, J. H. Booske, and N. Behdad are with the Department of  Electrical and Computer Engineering, University of Wisconsin–Madison, Madison, WI 53706 USA (e-mail: jzhao335@wisc.edu; jhbooske@wisc.edu; behdad@wisc.edu).

S. Qiao was with the Department of Electrical and Computer Engineering, University of Wisconsin-Madison, Madison, WI, 53706, USA  (email: shichen.qiao@berkeley.edu).
}}
\maketitle

\begin{abstract}
We present an electronically-reconfigurable antenna array offering low probability of intercept/detect (LPI/LPD) and secure communications capabilities simultaneously at the physical layer. This antenna array is designed to provide rapidly time-varying sidelobes and a stationary main lobe. By performing rapid sidelobe time modulation (SLTM), the signal transmitted in the undesired directions (i.e., through sidelobes) undergoes spread-spectrum distortion making it more difficult to be detected, intercepted, and deciphered while the signal transmitted in the desired direction (i.e., through the main lobe) is unaffected. Therefore, the intended receiver would not need additional modifications (i.e. encryption keys) to detect and recover the signal. We describe the operating principles of this SLTM array and validate its spread-spectrum SLTM sequence generation in undesired directions through theory, simulations, and experiments. Using a fabricated SLTM prototype operating at X band, we conducted system-level measurements to demonstrate its LPI/LPD, secure communications, and jamming resilience capabilities. The presented method is a physical layer technique, which can bring LPI/LPD capabilities to existing communications systems by simply replacing their antennas with SLTM arrays. This technique can be used independently or in combination with additional coding and signal-processing techniques to achieve further enhancements in LPI/LPD and secure communications.
\end{abstract}


%

\section{Introduction}
Wireless communication systems are widely used in civilian and military applications, providing reliable connectivity in various spheres of modern life. This reliance on readily accessible wireless communications systems is expected to continue with the development of 5G communications and the Internet of Things (IoT). However, transmission security and privacy are key concerns in many of civilian and military applications \cite{z1,z2,z3}. In most wireless communications systems, it is very important that the information contained in the transmitted signal can only be detected and deciphered by the intended recipient. In some applications, particularly military and security applications, hiding even the existence of a signal or transmission from unintended receivers is desirable. In such systems, even determining a transmission of interest has occurred by eavesdroppers can give the location of the transmitter away, which may have severe adverse consequences. On the flip side, it is always desired to have a clear receiver channel between the two desired nodes of communications and eliminate any jamming and interference from any unintended directions. To address these problems, the development of LPI/LPD secure communications systems is of particular interest.

{In \cite{z2,z3}, authors distinguish between LPI/LPD and secure communications using user objectives. Within LPI/LPD, eavesdroppers focus on detecting unknown transmitters. Conversely, in secure communications, their primary objective is decoding intercepted information. Typically, eavesdropper detection systems first identify the presence of the transmitter, then attempt to decipher the intercepted signal. Therefore, LPI/LPD secure communication systems should evade detection and deciphering by eavesdroppers to ensure covert communication from transmitters to receivers.}

Techniques which reduce the signal-to-noise ratio (SNR) or distort the signal detected by the eavesdroppers can enhance the LPI/LPD and physical-layer security of the system. These techniques include waveform design, artificial jamming, and antenna modification. Waveform design is a particularly effective method used in LPI/LPD \cite{stove_2004,Garcia_2000,Liu_2003,z4,z5,z6,z7} and secure communications\cite{z12,z13}. {Waveforms like frequency modulated continuous wave \cite{stove_2004}, frequency hopping\cite{Garcia_2000}, random noise\cite{Liu_2003}, spread spectrum\cite{z4,z6,z12,z13} and hybrid waveforms are designed to spread the energy of the signal in time or in frequency over a wide bandwidth, effectively reducing the SNR observed by eavesdroppers and distorting the signal.} However, these waveforms usually require additional processes at the intended receiver to detect or recover the signal, which increases the system complexity.

In \cite{z9,z10}, artificial noise is generated by the transmitter or a supplementary jammer. Adding artificial noise distorts the channel between the transmitter and the eavesdropper, reducing the SNR, and making it difficult for the eavesdropper to distinguish the signal from their noise floor. Since generating artificial noise requires extra jamming transmitters and/or an extra antenna, this will increase the complexity of the entire communication system. Additionally, leakage of the artificial noise from the jammer will cause undesired interference on the intended receiver. 

Antenna modification is another technique to address the problem. Refs. \cite{z20, C_1993, z21} presented the results of an ultralow sidelobe antenna achieved by controlling the phase and amplitude response of each array element in a conventional phased array. Ultra-low sidelobe beamforming can improve the LPI/LPD and security performance by reducing the SNR in the sidelobe direction. {However, this method only reduces signal power without altering its spectral characteristics or information content, leaving the communication system vulnerable to detection and deciphering, especially if eavesdropper is close to the transmitter and uses a sensitive detector.} At the same time, array fabrication error tolerances, both systematic and random, will degrade the sidelobe level (SLL) from the designed values \cite{z21}. 

{Time-modulated (TM) arrays, which change the array response with time, have also been explored to achieve enhancing LPI/LPD \cite{z16,z17,z19,z23,z24,chen_2020,Chen2_2022} and secure communications \cite{z16,z17,z19,z23,Rocca_2014,z25,sun_2018,Venkatesh_2021,Chen1_2022,Li_2022}.} By using a high-speed RF switch in each array element, TM arrays can synthesize ultra-low SLLs to reduce the SNR in the sidelobe directions \cite{z16,z17,z19,z23}, thus enhancing LPI/LPD and secure communications simultaneously. TM arrays generate high-order frequency components in the frequency domain when they periodically repeat the TM sequence, causing {sideband radiation (SR)}. SR spreads power into unwanted frequency bands, reducing antenna gain in the intended direction \cite{z23}. However, by smartly controlling the TM array, the SR can occur in unintended directions but not in the intended direction. {Therefore, this technique takes advantages of the spectrum aliasing effect, overlapping the spectrum in the center frequency with that in the sideband, thus distorting the signal for secure communications\cite{Rocca_2014,z25,sun_2018,Venkatesh_2021}. Ref. \cite{Chen1_2022} proposed a pseudorandom TM sequence in the undesired directions for increasing the randomness in secure communications. These methods, including periodic and pseudorandom TM sequences, significantly improved communication security, although LPI/LPD performance was not significant since the PSD around the center frequency remains prominent compared to the PSD outside the center frequency. In \cite{Li_2022}, chaotic sequences were used in the TM array to encrypt the transmitted signal, making it difficult for eavesdroppers to decipher the signal. Only the intended receiver knowing the chaotic sequence can decipher the signal. To achieve LPI/LPD, the radar signal was distorted by using various pseudorandom TM sequences to effectively hide it from eavesdroppers \cite{z24,chen_2020}. However, the SNR was not reduced and even increased in some unintended directions \cite{z24}, potentially making it easier for eavesdroppers with sensitive detectors to detect the signal. Furthermore, the receiver in \cite{chen_2020} needs a matched filter to recover the received signal, which requires knowledge of the transmitted waveform. While most previous studies focused on improving LPI/LPD or secure communications capabilities separately, we propose an effective TM-array-based approach to achieve both of these traits simultaneously. To do this, we present the concept of sidelobe time modulation (SLTM) in this paper. Important features of the SLTM concept and the unique contributions of this work over the state-of-the-art are:}
\begin{enumerate}
  \item {SLTM arrays provide extremely-fast beam manipulation in the sidelobe directions without impacting the TM-array's pattern in the main lobe direction.} 

  \item {We propose the SLTM sequence and its optimization procedure to scramble the radiated signal in an spread-spectrum-like manner in the sidelobe directions only. In these directions, the signal will be irrecoverably scrambled and its energy will be spread over a large bandwidth, thereby enhancing LPI/LPD and communications security simultaneously. Meanwhile, reciprocity indicates that SLTM arrays are less susceptible to sidelobe jamming. }

  \item { SLTM arrays can be employed in existing communications systems by simply replacing the existing antennas with SLTM arrays, without making additional system-level modifications. The intended receiver would not require additional signal processing techniques to detect or decipher the signal, since the main lobe is unaltered.}
\end{enumerate}
 
\makeatletter
\newcommand{\rmnum}[1]{\mathrm{\@Roman{#1}}}
\makeatother

The rest of the paper is organized as follows. {Section $\rmnum{2}$ explains the working principles and design of the SLTM sequence. Section $\rmnum{3}$ describes the design of an eight-element SLTM antenna array, including simulation results and measurements of radiation patterns.} In Section $\rmnum{4}$, we present results of numerical simulation based analyses of the LPI/LPD and secure communication performance of the SLTM antenna array, including PSDs and bit error rates (BERs). Section $\rmnum{5}$ presents measurement PSD and BER characteristics that validate the numerical analysis of the proposed method.  Finally, Section $\rmnum{6}$ provides conclusions.

\section{Principles of Operation of the Sidelobe Time-modulated Antenna Array}
\label{section2}

When the antenna works as a transmitter, eavesdropper's detection systems, often referred to as Radar Warning Receivers or Electronic Support Measures, work by detecting, identifying, and processing signals from transmitter systems. These systems are constantly scanning for RF emissions across a wide range of frequencies. When a transmitted signal is detected, it is because the output power from the transmitter is above the detection threshold of the eavesdropper. Once a signal is detected, the system analyzes the characteristics of the signal, such as its PSD, pulse width, pulse repetition frequency, and other properties. These characteristics can be used to identify the type of transmitter. After identification, the system can evaluate the level of threat and proceed with signal processing. If the detected transmitter is determined to be a threat, the system can trigger a response or an alarm. The goal of an LPI/LPD strategy is to undermine these procedures and enable detection- and decipher-proof transmission and jamming-proof reception.

\begin{figure}[h]
\centering
\includegraphics[width=0.47\textwidth]{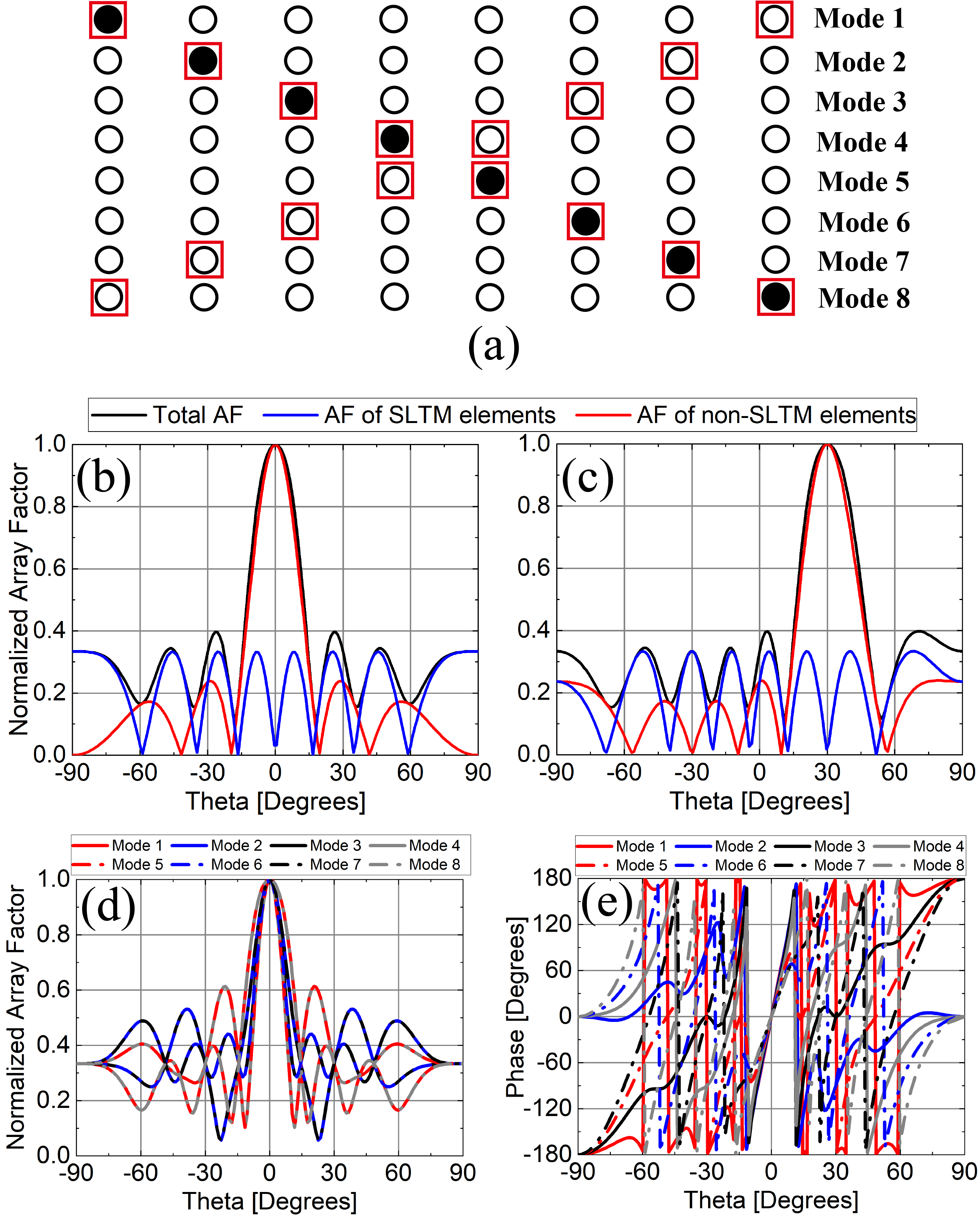}
\caption{(a) Eight operational modes of an eight-element SLTM array. (b) Array factor of SLTM elements, array factor of non-SLTM elements, and their total array factor in 0$^{\circ}$ in Mode 1. {(c) Array factor of SLTM elements, array factor of non-SLTM elements, and their total array factor in 30$^{\circ}$ in Mode 1. (d) Amplitude of array factor in 8 different modes in 0$^\circ$. (e) Phase of array factor in 8 different modes in 0$^\circ$. (All the figures are generated when $d$ is a half wavelength)}
\label{SLTM_matlab}}
\end{figure}

To demonstrate the concept, let us consider a one-dimensional, linear SLTM array. The signal transmitted from this SLTM array can be expressed as 
\begin{equation}
y(\theta,t)=x(t) \cdot E(\theta,t)
\label{eqn0}
\end{equation}
where $x(t)$ is the time-domain representation of the modulated RF signal to be transmitted and $E(\theta,t)$ is the electric field of the SLTM array. To be compatible with the time-dependent complex format of the array factor, $x(t)$ is a complex signal. Assuming that all array elements are identical, the time-dependent electric field of this antenna array  can be expressed as:
\begin{equation}
\centering
E(\theta,t) =e_0(\theta)[\sum\limits_{k=1}^{N} G_k(t)\cdot I_ke^{j\alpha_k}\cdot e^{j\beta(k-1)d\sin(\theta)}]
\label{equ1} 
\end{equation}
{where $e_0(\theta)$ is the element pattern}, $G_k(t)$ represents the time-modulated phase state coefficient ($G_k$=+1 or -1), $I_k$ and $\alpha_k$ are the initial amplitude and phase of the $k^{th}$ element, $\beta$ is the wavenumber in free space, $d$ is the space between two adjacent elements (which we assume is uniform), and $\theta$ is the observation angle with reference to the broadside direction. {To simplify the analysis, we have assumed that all elements are linearly-polarized and that the $E(\theta,t)$ in (\ref{eqn0})-(\ref{equ1}) represents the co-polarized component of the electric field in the far-field region of the antenna in the plane containing the array.} The portion inside the brackets of (\ref{equ1}) signifies the TM array factor, achieved by switching the value of $G_k$ between +1 and -1. As depicted in Fig. \ref{SLTM_matlab}(a), the white circles represent isotropic array elements excited in phase ($G_k$=1), whereas the black circles symbolize array elements excited out of phase ($G_k$=-1). One possible strategy for generating radiation patterns used in SLTM is to sequentially invert the phase of a single-element in the array as shown in Fig. \ref{SLTM_matlab}(a). This way, eight different operating modes are produced, each with its own radiation pattern. We can think of the pairs of out-of-phase array elements (highlighted in red boxes in Fig. \ref{SLTM_matlab}(a)) as the elements used in SLTM and the in-phase elements as the ones forming the main beam. As illustrated in Fig. \ref{SLTM_matlab}(b), the array factor of the paired SLTM elements cancels out in the broadside direction and only impacts the undesired directions. {Fig. \ref{SLTM_matlab}(c) shows that the same concept applies when the main beam is steered off-broadside by adjusting the initial phase values, $\alpha_k$. In this case, the array factors of both the SLTM and non-SLTM elements are steered towards 30$^\circ$.} In principle, this should allow one to change the total array factor in undesired directions while not significantly impacting the array factor along the main lobe direction {whether the main lobe points towards broadside or off-broadside directions.} In our linear array example discussed here, there are eight modes, i.e., eight sets of radiation patterns with unique magnitudes and phases as presented in Fig \ref{SLTM_matlab}(d) and Fig. \ref{SLTM_matlab}(e). By pseudorandomly and rapidly changing the radiation pattern of this SLTM array, an additional time modulation is applied to $x(t)$ to produce the new TM waveform. By properly designing this waveform, we can obtain LPI/LPD operation and resilience to jamming/interference.
   
For convenience, we can set the reference phase of the main lobe to $0^{\circ}$ corresponding to $\alpha_k$=0 {and assume that the array elements are isotropic (or onmidirectional in the observation plane) which results in $e_0(\theta)$=1. In this case, (\ref{equ1}) can be expressed as:} 
\begin{equation}
\centering
\begin{aligned}
E(0^{\circ},t) &={\sum\limits_{k=1}^{N} G_k(t) = C_0 \cdot e^{j\phi_{c}},} &\text{In\ main\ lobe }\\
E(\theta_0,t) &={\sum\limits_{k=1}^{N} G_k(t)I_ke^{j\beta(k-1)d\sin(\theta_0)},} &\text{In\ sidelobe} \\
                  &= C(\theta_0,t)e^{j\phi(\theta_0,t)} 
\label{equ2}
\end{aligned}
\end{equation}
In the main lobe, the summation of $G_k$ is always equal to 6 when the SLTM array is switching between these eight modes. Therefore, the array factor in the main lobe is unchanged with a constant amplitude, $C_0$, and phase, $\phi_c$. In the sidelobe, the array factor changes in time with amplitude $C(\theta_0,t)$ and phase $\phi(\theta_0,t)$. In the sidelobe direction $\theta_0$, $C(\theta_0,t)e^{j\phi(\theta_0,t)}$ has eight different values. By continuously and randomly changing the array factor with time modulation frequency $f_{SLTM}$, we can obtain the TM sequence with the bit length of M. Here, M is an arbitrary number, which is determined by the user. The TM sequence is expressed as:
\begin{equation}
\begin{aligned}
C(\theta_0,t)e^{j\phi(\theta_0,t)} =\sum\limits_{m=0}^{M} c_m \cdot w(t-mT_{SLTM}) \\
\label{equ3}
\end{aligned}
\end{equation}
In these expressions, $c_m$ is a discrete random variable taking distinct complex values, $C_ie^{j\phi_i}$, to represent the coefficient of the $m^{th}$ element in the TM sequence. It follows a generalized Bernoulli distribution and its probability mass function can be expressed as:
\begin{subequations}
\begin{equation}
P(c_m=C_ie^{j\phi_i})=p_i\\
\label{}
\end{equation}
\begin{equation}
\sum\limits_{i=1}^{K}p_i=1
\label{}
\end{equation}
\label{pmf1}
\end{subequations}
where $p_i$ is the probability of taking the element value of $C_ie^{j\phi_i}$. $C_i$ and $\phi_i$ denote the amplitude and phase of $E(\theta_0,t)$ when the array works in Mode i ($i = 1, 2, ..., 8$). Therefore, K=8 in the illustrative example investigated in this paper. $w(t)$ is the window function which satisfies (\ref{wt}),
\begin{equation}
w(t)=\left\{
\begin{aligned}
&1,\quad  0<t \leq T_{SLTM}  \\
&0,\quad  t > T_{SLTM},
\label{wt}
\end{aligned}
\right.
\end{equation} 
and $T_{SLTM} = \frac{1}{f_{SLTM}}$ is the duration of the sequence element. 

Assuming that the bit duration of $x(t)$ is $T_b$ in (\ref{eqn0}), we assume that $T_{SLTM} \ll T_b$ since the SLTM array is rapidly changing the radiation pattern and applying the TM sequence $E(\theta_0,t)$ on $x(t)$. Therefore, in the frequency domain, the PSD of $y(\theta_0,t)$ is equal to the convolution of a narrow band PSD of $x(t)$ and a wideband PSD of $E(\theta_0,t)$. This wideband PSD of the TM sequence will significantly change the characteristics of the transmitted signal $y(\theta_0,t)$ in the frequency domain. For example, if $x(t)$ is a harmonic, single-frequency CW signal whose PSD in the frequency domain is a $\delta$ function, the PSD of $y(\theta_0,t)$ is equal to the PSD of $E(\theta_0,t)$ convolved with the $\delta$ function. In this case, we focus on analyzing the PSD of $E(\theta_0,t)$ and designing the waveform of $E(\theta_0,t)$ that can oabtain LPI/LPD secure communication and jamming resilience.


To analyze the PSD of $E(\theta_0,t)$, we decomposed $E(\theta_0,t)$ into a real part, $r(t)$, and an imaginary part, $s(t)$, in the sidelobe direction $\theta_0$:
\begin{equation}
\centering
\begin{aligned}
E(\theta_0,t)&=\text{Re}\{E(\theta_0,t)\} + j\text{Im}\{E(\theta_0,t)\} \\
        &= r(t) + js(t).
\label{equ6}
\end{aligned}
\end{equation}
The PSD of $E(\theta_0,t)$ is equal to the sum of the PSD of the real part and the PSD of the imaginary part. Here, we only analyze the PSD of the real part since the analysis procedure of the PSD of the imaginary part is the same as that of the PSD of the real part.


From (\ref{equ3}), we can obtain the real part of the TM sequence, $r(t)$, comprising M symbols:
\begin{equation}
r(t) =\sum\limits_{m=0}^{M} A_m \cdot w(t-mT_{SLTM})=\sum\limits_{m=0}^{M}r_m(t)
\label{rt}
\end{equation}
$r_m(t)$ is the $m^{th}$ symbol within $r(t)$. $A_m$ is equal to $Re[c_m]$ which is also a discrete random variable taking distinct values $a_i$. The amplitude of the real part of the TM sequence is given by $a_i$ ($i$=1, 2, ... 8) for an 8-element array. The value of $a_i$ changes with variations in the radiation pattern and satisfies $a_i=Re\{C_ie^{j\phi_i}\}$ with different probabilities $p_i$. The probability mass function of $A_m$ can be expressed as:
\begin{equation}
\begin{aligned}
P(A_m=a_i)=p_i
\label{pmf2}
\end{aligned}
\end{equation}
Then we decompose stochastic signal $r(t)$ into two components to analyze its PSD 
\begin{equation}
r(t) =v(t)+u(t) 
\end{equation}
where $v(t)$, the statistical average of $r(t)$, is equal to the expectation of $r(t)$ in each duration $T_{SLTM}$ and can be calculated as follows. As shown in (\ref{vt}), $v(t)$ can be expressed as a DC signal with constant amplitude $[p_1a_1+p_2a_2+\cdots +p_ia_i]$ in the time domain. This amplitude is the expectation of $A_m$ and can be expressed as {$\E[A_m]$}:
\begin{equation}
\begin{aligned}
v(t)&=\sum\limits_{m=0}^{M} [p_1a_1+p_2a_2+\cdots +p_ia_i] \cdot w(t-mT_s)\\
&=\sum\limits_{m=0}^{M}{\E[A_m]} \cdot w(t-mT_s)=\sum\limits_{m=0}^{M} v_m(t).
\end{aligned}
\label{vt}
\end{equation}
Meanwhile, $u(t)$, the stationary random process of $r(t)$, is the difference between $r(t)$ and $v(t)$:
\begin{equation}
\begin{aligned}
u(t) &=r(t)-v(t)\\
 \sum\limits_{m=0}^{M} u_m(t)   &=\sum\limits_{m=0}^{M} r_m(t)-\sum\limits_{m=0}^{M} v_m(t)
\end{aligned}
\label{ut}
\end{equation}
So the $m^{th}$ symbol of $u(t)$ can be expressed by substituting (\ref{rt}) and (\ref{vt}) into (\ref{ut}):
\begin{equation}
\begin{aligned}
u(t) &=\sum\limits_{m=0}^{M} (A_m-{\E[A_m]}) \cdot w(t-mT_{SLTM})\\
      &= \sum\limits_{m=0}^{M} B_m \cdot w(t-mT_{SLTM}).
\label{ut3}
\end{aligned}
\end{equation}
Here, $B_m$ is the amplitude of each symbol in $u(t)$ which is equal to $A_m-{\E[A_m]}$. Since $A_m$ is a discrete random variable, $B_m$ is determined by the value of $A_m$. We use $b_i$ to represent the possible value that $B_m$ is taking:
\begin{equation}
\resizebox{.85\hsize}{!}{$B_m=\left\{
\begin{aligned}
b_1&= (1-p_1)a_1-p_2a_2-\cdots -p_ia_i , & \ p_1\\
b_2&=-p_1a_1  +(1-p_2)a_2-\cdots -p_ia_i, & \ p_2\\
 & \qquad \qquad \qquad \vdots & \\
 b_i&=-p_1a_1-p_2a_2-\cdots +(1-p_i)a_i, & \ p_i
\end{aligned}
\right.$}
\label{ut2}
\end{equation}

The PSDs of $v(t)$ and $u(t)$ are derived individually and then combined to obtain the PSD of the SLTM sequence. Based on the conclusion from the Fourier transform, if $v(t)$ is a periodic signal, its PSD can be found by computing the Fourier series of $v(t)$.  As mentioned, $v(t)$ is a periodic DC signal, thus, its Fourier series only has the DC impulse component with the amplitude of {$\E[A_m]$}. The PSD of the periodic signal is equal to the discrete $\delta$ function multiplied by the square of its Fourier series. Therefore, the PSD of $v(t)$ can be expressed as:
\begin{equation}
\begin{aligned}
P_{rv}(f)&=|(p_1a_1+p_2a_2+
\cdots+p_ia_i)|^{2}\delta(f) \\
&=|{\E[A_m]}|^{2}\delta(f)
\end{aligned}
\label{eqn8a}
\end{equation}
The PSD of the stationary random process $u(t)$ can be deduced by evaluating its statistical average in the time domain. The derivation procedure is detailed in Appendix and will not be repeated here. Finally, the PSD of the stationary random process $P_{ru}(f)$, and the PSD of the real part of the SLTM sequence $P_r(f)$ can be calculated as:
\begin{subequations}
\begin{equation}
P_{ru}(f)=\left(\frac{{\E[B_{m}^{2}]}}{f_{SLTM}}\right)|\sinc(\pi fT_{SLTM})|^{2}
\label{eqn8b}
\end{equation}
\begin{equation}
P_r(f)=P_{rv}(f)+P_{ru}(f).
\label{eqn8c}
\end{equation}
\label{eqn8}
\end{subequations}
Here, {$\E[B_{m}^{2}]$} is the expectation of  $B_m^{2}$ (cf.(\ref{ut2})) which can be expressed as below:
\begin{equation}
\centering
\begin{aligned}
{\E[B_{m}^{2}]}&= \sum\limits_{i}^{}b_{i}^{2}p_i \\
                 &= \sum\limits_{i}^{}a_{i}^{2}p_i(1-p_i) - 2\sum\limits_{i}^{}\sum\limits_{q \neq i}^{}a_ia_qp_ip_q
\label{bi}
\end{aligned}
\end{equation}
and $sinc(f)$ is the $sinc$ function, which is the Fourier transform of the window function, $w(t)$. The PSD of $v(t)$ appears as a discrete DC impulse, whereas the PSD of $u(t)$ is a continuous $sinc$ function.

\begin{figure}[h]
\centering
\includegraphics[width=0.45\textwidth]{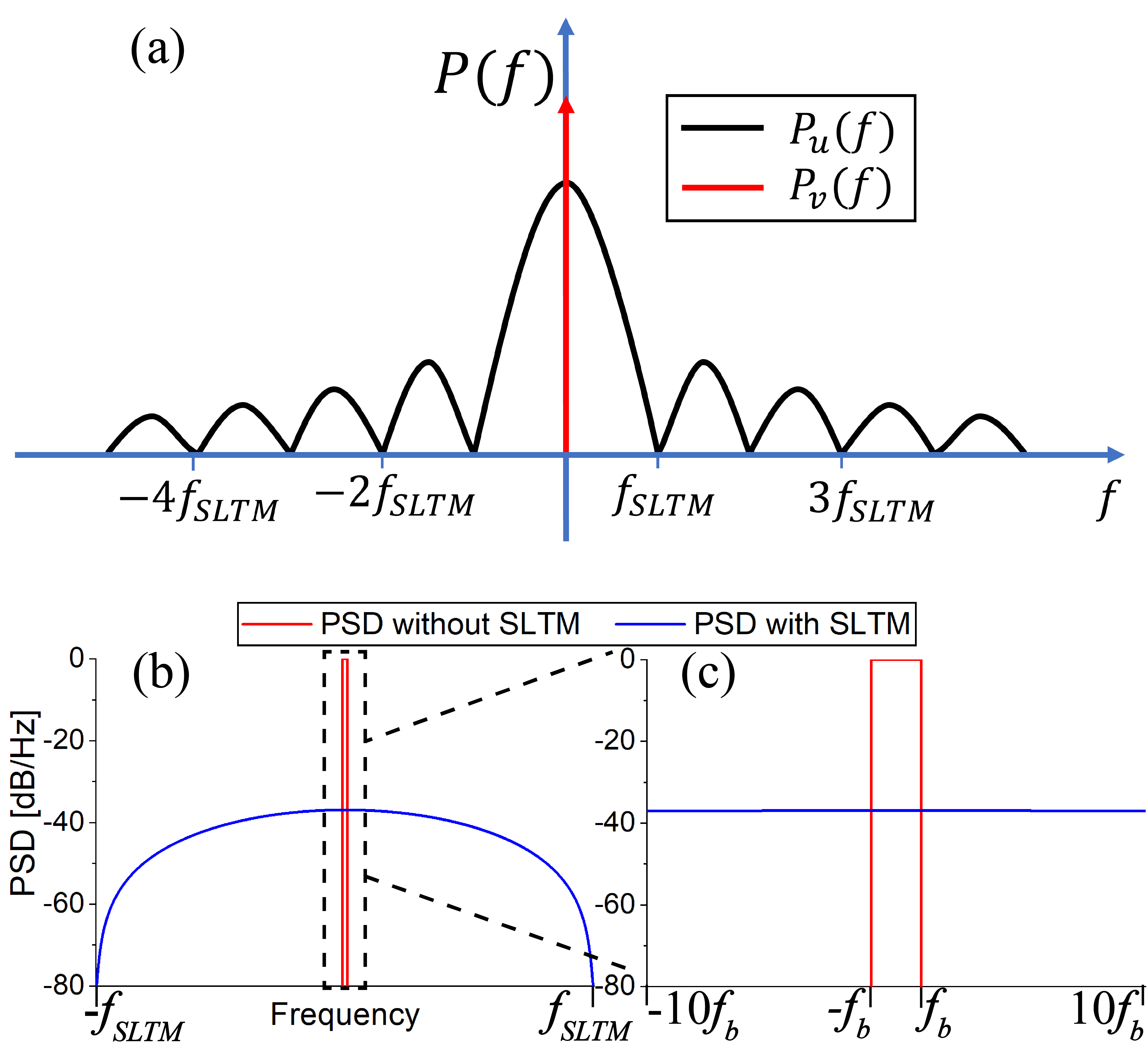}
\caption{ {Graphical representation of the power spectral density of the (a) SLTM sequence, (b) transmitted signal in main lobe and sidelobe directions and (c) zoom-in view of the transmitted signal. $f_b$=$\frac{1}{T_{b}}$.}
\label{fig3}}
\end{figure}

As mentioned, the PSD of the imaginary part, $s(t)$, is in the same form as (\ref{eqn8a}) and (\ref{eqn8}), with the only difference being $a_i=Im\{C_ie^{j\phi_i}\}$ in (\ref{pmf2}). Therefore, the PSD of the TM sequence, $P(f)$, is:
\begin{equation}
P(f)=P_{r}(f)+P_{s}(f)=P_{rv}(f)+P_{ru}(f)+P_{sv}(f)+P_{su}(f)
\label{eqn_Pf}
\end{equation}
where $P_{s}(f)$ is the PSD of the imaginary part, $P_{sv}(f)$ is the PSD of the statistical average component in $P_{s}(f)$, and $P_{su}(f)$ is the PSD of the stationary random process in $P_{s}(f)$. The PSD of the TM sequence, $P(f)$, is depicted in Fig. \ref{fig3}(a), with $P_{v}(f)=P_{rv}(f)+P_{sv}(f)$ and $P_{u}(f)=P_{ru}(f)+P_{su}(f)$.

\begin{figure}[h]
\centering
\includegraphics[width=8cm]{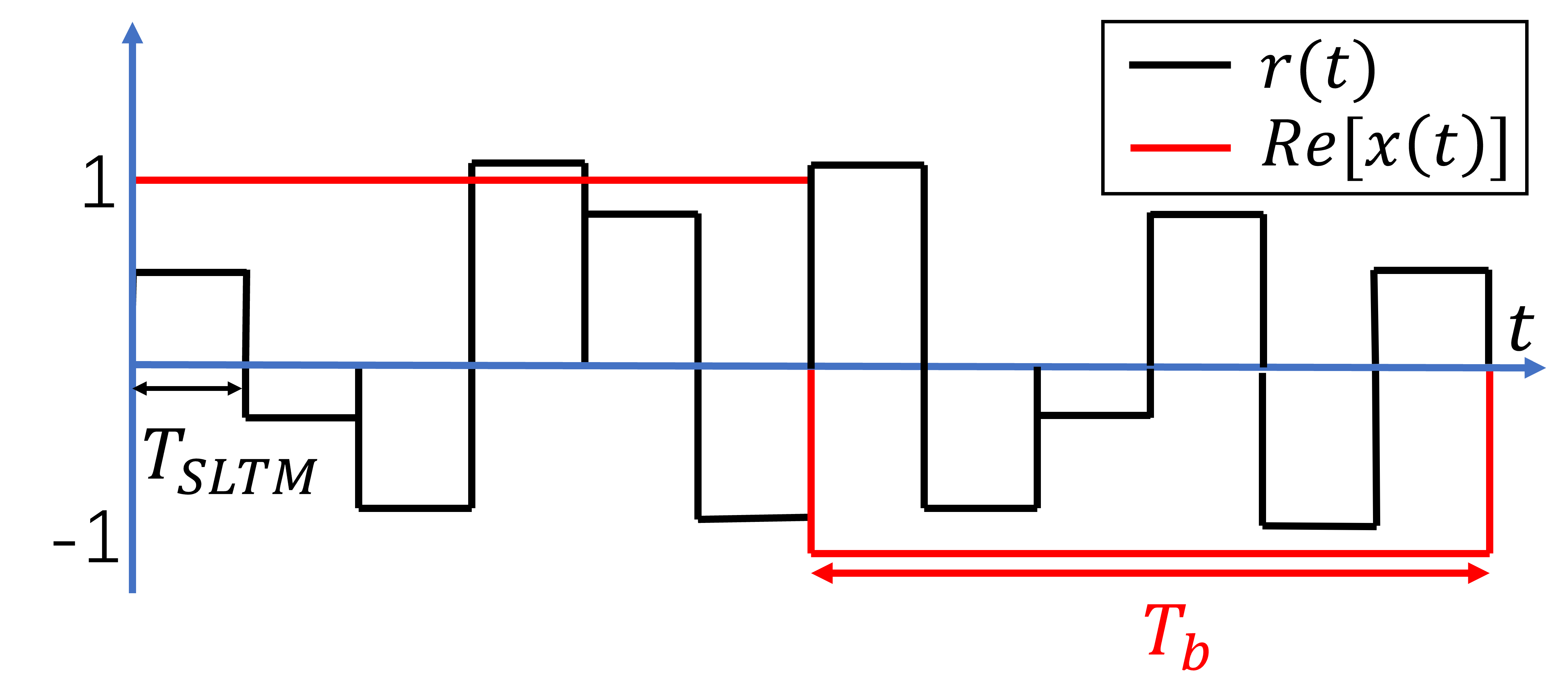}
\caption{ Graphical representation of the time domain $Re[x(t)]$ and the real part of the TM signal $r(t)$.
\label{fig4}}
\end{figure}

From (\ref{eqn8a}), it can be observed that the amplitude of the $\delta$ function at DC is determined by the expectation of $A_m$ given by the expression $p_1a_1+p_2a_2+\cdots +p_ia_i$. {(\ref{eqn8b}) indicates that the null-to-null bandwidth of the TM signal, $E(\theta_0,t)$, extends up to $2f_{SLTM}$, spreading the power from the narrow band of $x(t)$ to the wider band of the TM signal as shown in Figs. \ref{fig3}(b) and \ref{fig3}(c). In these figures, the red curve represents the normalized PSD of $y(\theta,t)$ in the main lobe direction while the blue curve is the PSD in the sidelobe direction. These curves are plotted based on the assumption that $B_m^{2}$=0.01 and $f_{SLTM}$=100$f_b$, which are reasonable in an SLTM array.} The amplitude of $P_{ru}(f)$ is directly proportional to the expectation of $B_m^{2}$ and inversely proportional to the SLTM frequency $f_{SLTM}$. To enable the SLTM system to distort the signal and hide its presence from eavesdroppers, the SLTM waveform must be optimized according to the analyses of (\ref{eqn8a}) and (\ref{eqn8}). One critical aspect to note is that the presence of the $\delta$ function in (\ref{eqn8a}) is undesirable. If we convolve the PSD of $x(t)$ with $E(\theta_0,t)$, the PSD of $x(t)$ will convolve with this $\delta$ function, resulting in the PSD of $y(\theta_0,t)$ containing the entire PSD of $x(t)$ with only an amplitude change as one of the components of the PSD. Therefore, $x(t)$ is successfully transformed to $y(\theta_0,t)$, meaning that the waveform characteristics and signal information could be retrieved. This $\delta$ function component, when present, not only enhances the PSD of the signal transmitted through the sidelobes but also makes the information within this signal more susceptible to interception and decoding by potential eavesdroppers. We can eliminate the $\delta$ function by ensuring {$\E[A_m]=0$} for both $r(t)$ and $s(t)$. Since $A_m= Re[E(\theta_0,t)]$ or $Im[E(\theta_0,t)]$, we can make {$\E[E(\theta_0,t)]=0$} to satisfy $\E[A_m]=0$ for both $r(t)$ and $s(t)$. This is important since using one variable reduces the optimization complexity compared to using two variables. We will elaborate on the optimization procedure in the following sections of the paper. In this scenario, the SLTM waveform is produced by selecting a proper combination of eight distinct modes. The eight modes are pseudorandomly selected to construct a TM sequence. The probabilities of each mode are selected in a manner to ensure the elimination of the $\delta$ function. Once the $\delta$ function is removed, $P_{rv}(f)=P_{sv}(f)=0$, and (\ref{eqn8c}) can be simplified to $P(f)=P_{ru}(f)+P_{su}(f)$. As inferred from (\ref{eqn8b}), increasing the SLTM frequency $f_{SLTM}$ can broaden the bandwidth of the transmitted signal across a wider frequency band, thereby improving the LPI/LPD and protecting the embedded information. Meanwhile, minimizing the expected value of $B_m^{2}$ can further reduce the PSD across the frequency band of interest. Fig. \ref{fig4} illustrates the time domain sequence of $r(t)$ and the real part of $x(t)$. In this figure, a rectangular wave is used to represent the real part of $x(t)$, but it can be any arbitrary waveform.

To conclude, the SLTM waveform should possess the following properties, listed in order of priority:
\begin{enumerate}
  \item The expectation of $A_m$ approaches  0.
  \item The SLTM frequency $f_{SLTM}$ is as high as possible compared to the bandwidth of $x(t)$.
  \item The expectation of $B_m^{2}$ approaches  0.
\end{enumerate}
In this paper, the SLTM waveform is optimized subject to the first and second properties to simplify the optimization procedure.

\section{SLTM Antenna Array Design}
\label{section3}
To create the aforementioned SLTM waveform, which maintains performance in the main lobe while scrambling in the sidelobe, an antenna array was developed that allows for rapidly modulating its radiation pattern. SLTM is achieved when the array rapidly changes the amplitude and phase in sidelobes without significantly impacting the main lobe. Furthermore, the array element should be capable of rapidly switching the modulation coefficient ($G_k$) between +1 and -1. In \cite{z28}, Min et al. developed a Ku-band 1-bit phase-reconfigurable patch element capable of generating a 1-bit ($0^{\circ}$/$180^{\circ}$) phase state. The feed probe is in the center of the patch and connected to the patch by two symmetrically embedded PIN diode switches. By controlling switches, the surface current flow is reversed to generate $180^{\circ}$ phase shift.  In this paper, by taking advantage of the topology of this 1-bit patch and simplifying its DC bias lines, we design, simulate, and measure an SLTM array operating at 10 GHz using the 1-bit patch as an array element.

\begin{figure}[h]
\centering
\includegraphics[width=0.47\textwidth]{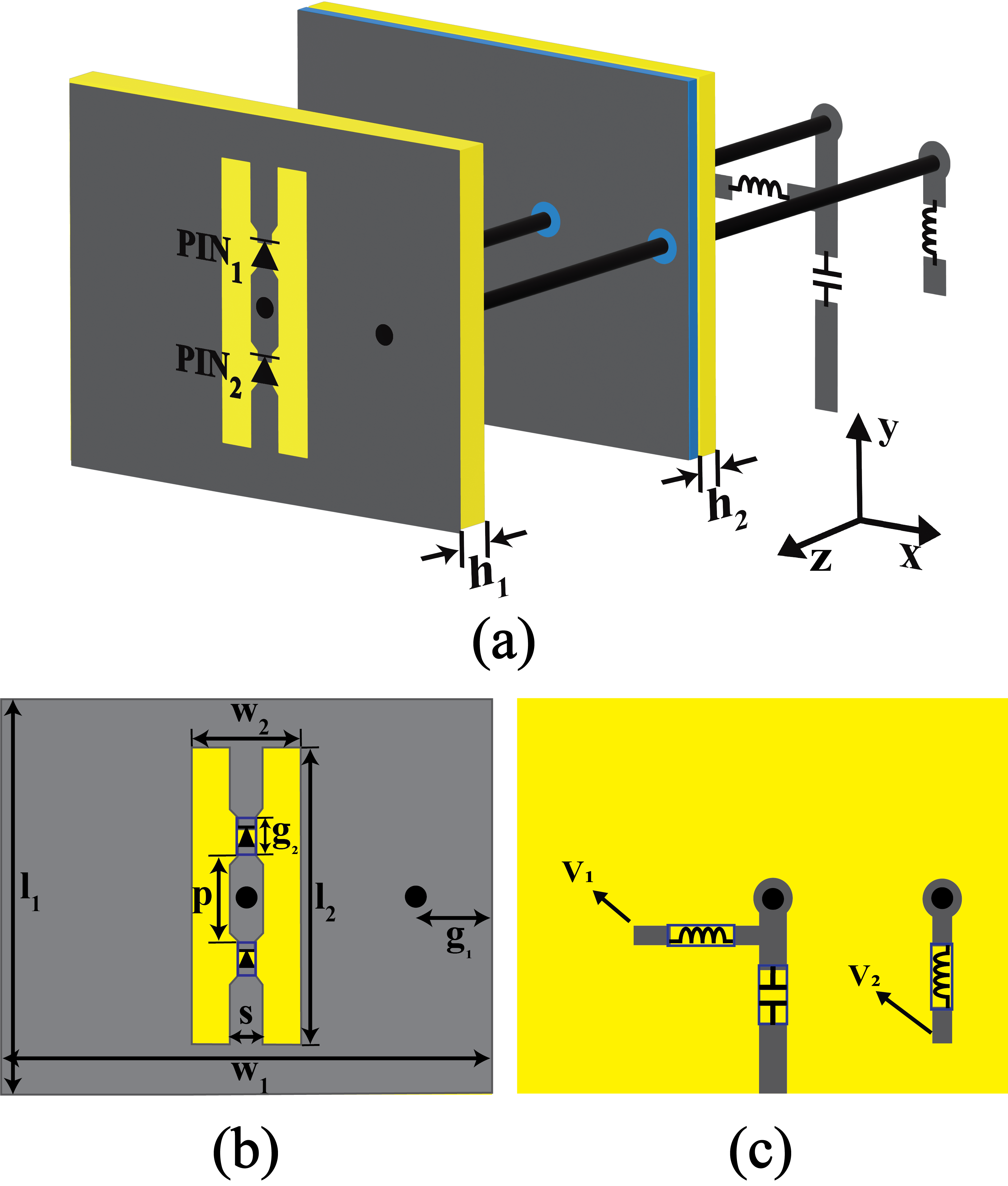}
\caption{Topology of the slotted patch antenna with a 1-bit electronically reconfigurable phase shift. (a) 3D perspective view. {(b) Top view}. (c) Bottom view.
\label{patch_model}}
\end{figure}

\subsection{One-bit, Phase-reconfigurable Antenna Element}

%

Fig. \ref{patch_model} depicts the topology of the 1-bit phase-reconfigurable patch antenna used in the construction of the SLTM antenna array. In this configuration, the patch substrate and feed line substrate are separated by a ground plane. Above the ground plane, the patch is supported by a single-layer of Rogers RO4003C. Beneath the ground plane, another layer of RO4003C substrate, adhered to the first one using a layer of Rogers RO4450F prepreg, serves to support the feed line. This patch consists of three metallic layers. The top layer features the slotted patch, integrated with two pin diodes. These diodes are symmetrically positioned within the center of the slot and are connected to each other using a conducting strip at the center of the slot. The ground plane layer, positioned in the middle, contains two circular slots through which two vias connect the antenna and feed line layers together. The bottom layer contains a microstrip feed line equipped with DC bias circuits for the PIN diodes. These bias circuits allow the center connection point of the two diodes to be connected to a DC bias voltage of V$_1$ and the body of the patch to be connected to a DC bias voltage of V$_2$. On the top layer, a slotted patch is printed. The feeding probe, positioned centrally on the patch, extends through the circular slot in the ground plane, connecting to the microstrip line on the bottom metallic layer. A DC bias via traverses from the patch to the DC bias line on the bottom layer, serving to bias the PIN diodes. This via is strategically placed where the surface current of the patch is minimized, thereby reducing the loading effect from the DC bias circuit  and enhancing the RF/DC isolation. {Both the via and the feeding probe have a diameter of 0.4 mm.} On the bottom layer, a 50 $\Omega$ microstrip feed line transmits the signal from the port to the feeding probe. The key parameters of the proposed structure, after tuning through full-wave simulations, are given in Table \ref{table1}. 

\begin{table}[h]
\renewcommand{\arraystretch}{1.3}
\caption{Dimensions for 1-bit Patch Antenna}
\label{table1}
\centering
\stackunder{
\begin{tabular}{|c|c|c|c|c|c|c|c|c|c|}
\hline
\bfseries $l_{1}$ & \bfseries $w_{1}$ & \bfseries $l_{2}$ & \bfseries $w_{2}$ & \bfseries $s$ & \bfseries $g_1$ & \bfseries ${g_{2}}$ & \bfseries ${p}$ & \bfseries $h_{1}$ & \bfseries $h_{2}$\\
\hline
7.0 & 9.0 & 5.3 & 2.0 & 0.6 &1.5& {0.65}& {1.5}& 0.8 & 0.3\\
\hline
\end{tabular}
}{\parbox{3in}{
\footnotesize All dimensions are in mm.}
}
\end{table}

\begin{figure}[h]
\centering
\includegraphics[width=0.35\textwidth]{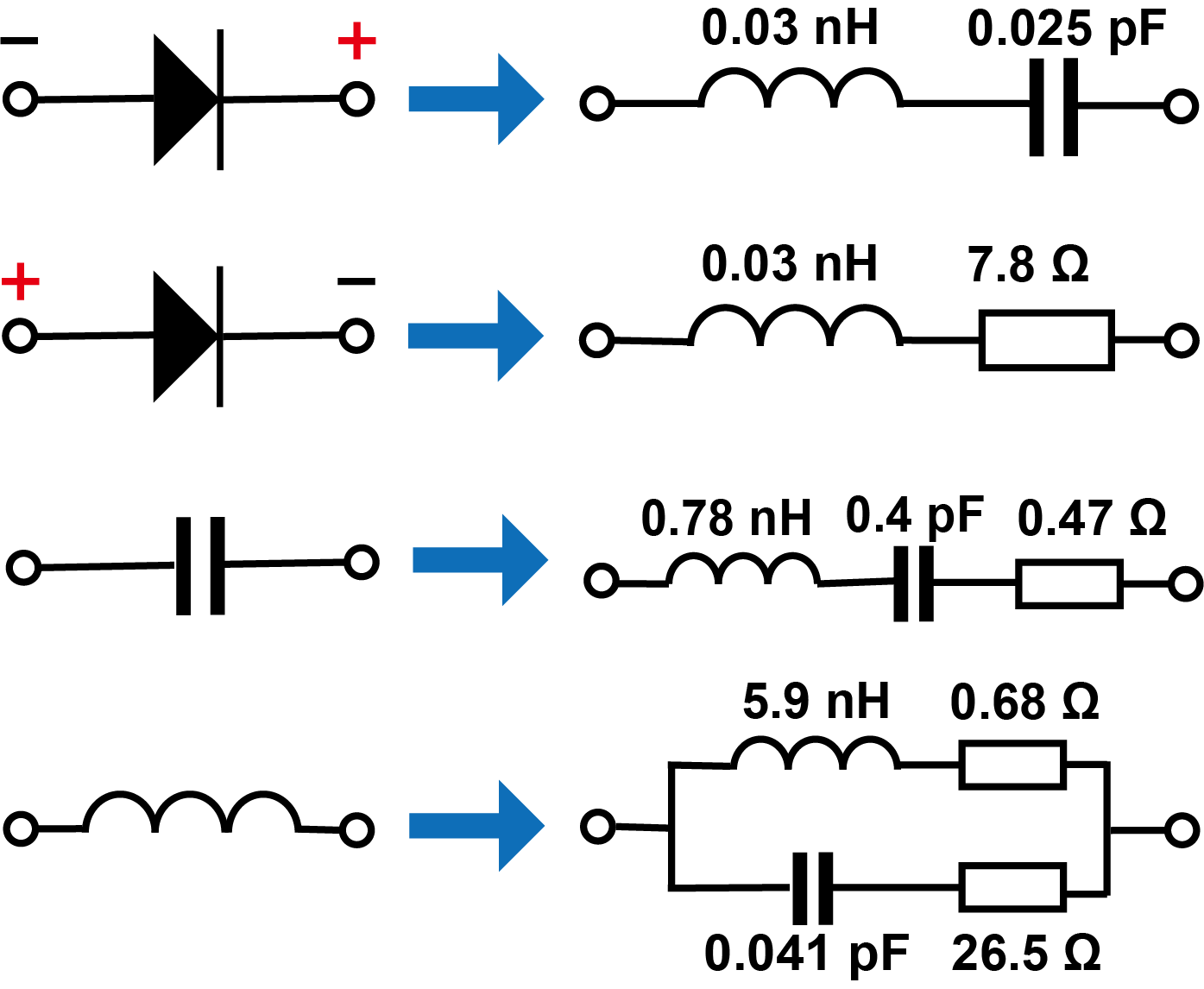}
\caption{Equivalent circuit models of the PIN diodes, capacitors and inductors employed in the design of the 1-bit, phase-reconfigurable patch antenna.
\label{fig_eq_circuit}}
\end{figure}

The PIN diodes are oriented in the same direction as shown in Fig. \ref{patch_model}(a). These diodes function as SPST switches, with their ON and OFF states controlled by forward and reverse biasing them, respectively. Due to their orientation, the two PIN diodes always have opposing bias states relative to each other. When one is ON, the other is OFF, and vice versa. In this design, we have incorporated PIN diodes manufactured by MACOM (model number MADP-000907) to ensure high-speed switching performance for the 1-bit patch. The switching speed of this PIN diode is in the order of 2 to 3 ns, which was the fastest commercially-available PIN diode that we could acquire at the time of fabrication. Two RF inductors (model: 0402DC-5N9 from Coilcraft) are integrated into each DC bias line to isolate the DC from the RF signal. Additionally, an RF capacitor (model: Accu-P 0603-0r4 from AVX) is incorporated into the microstrip feed line to block the DC signal. The equivalent circuits of the aforementioned components are modeled as lumped elements in the full-wave simulation. This approach takes the parasitic effects of the lumped components used in this design into account. Fig. \ref{fig_eq_circuit} shows the equivalent circuit models of the lumped element components used in the full-wave electromagnetic simulations. The same PIN diode model was characterized and used in several studies for 1-bit reconfigurable reflectarrays and its equivalent circuit model was shown to provide a good agreement with its measured response \cite{Luyen_2022,Yang_2016}. The values of the inductor and capacitor were adjusted to achieve a close alignment between the scattering parameters of the circuit model and the measured data provided by the manufacturers \cite{coilcraft,avx}.

\begin{figure}[h]
\centering
\includegraphics[width=0.47\textwidth]{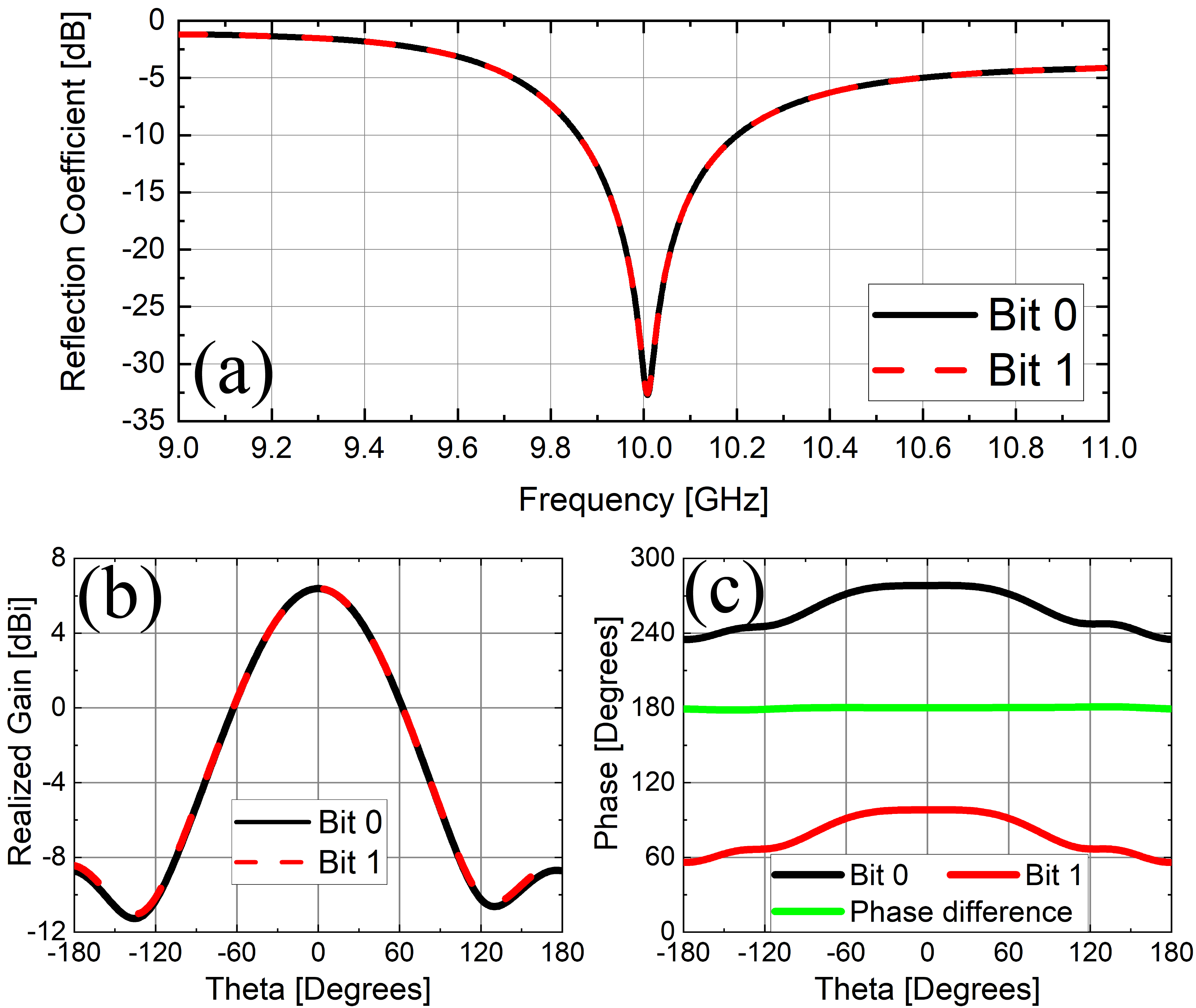}
\caption{(a) Simulated reflection coefficients of the 1-bit patch for bit-0 and bit-1. (b) Amplitude of the radiation pattern of the antenna for bit-0 and bit-1 at 10 GHz. (c) Phase of the radiation pattern of the antenna for bit-0 and bit-1 at 10 GHz.
\label{sim_patch}}
\end{figure}

%

\begin{figure}[h]
\centering
\includegraphics[width=0.5\textwidth]{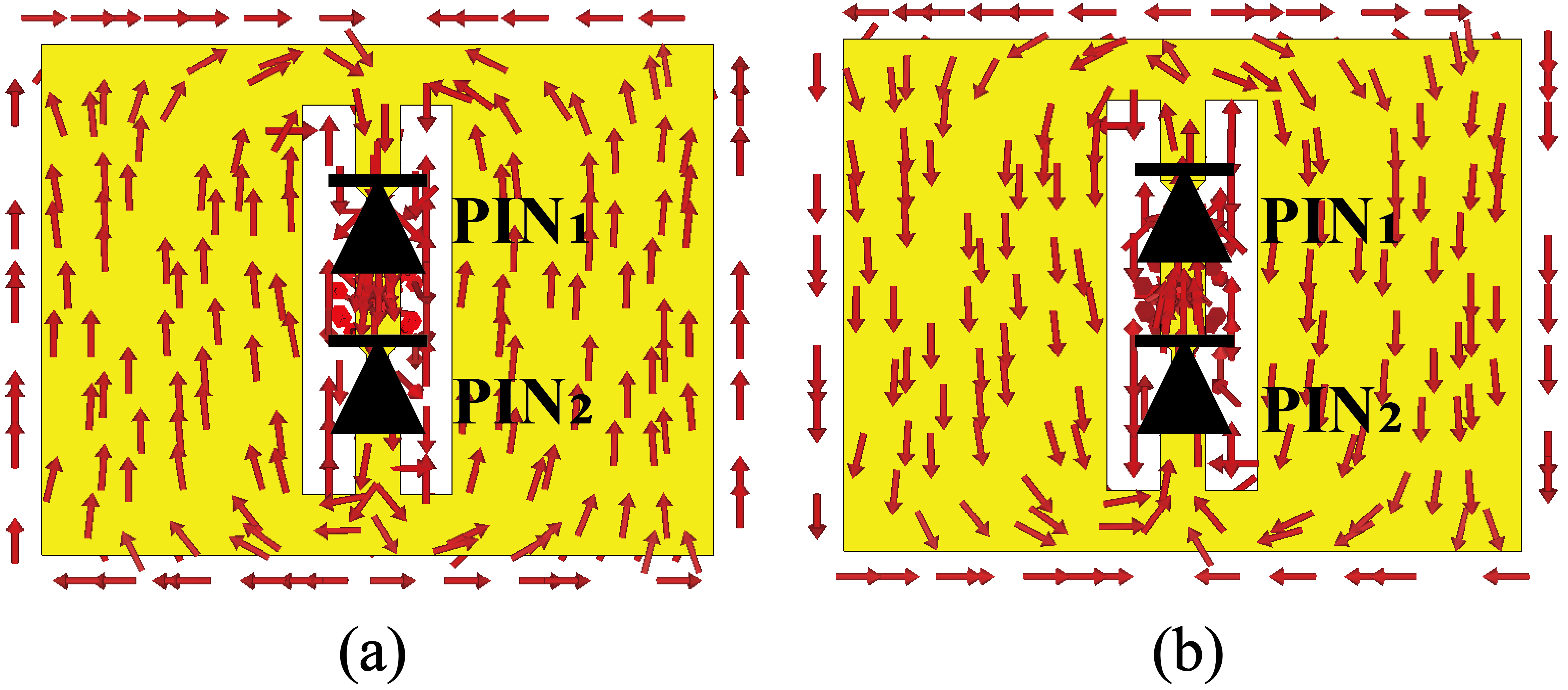}
\caption{ (a) Surface current flow of the antenna for bit-0. (b) Surface current flow of the antenna for bit-1.}
        \label{current_flow}
\end{figure}

We simulated this 1-bit, phase-reconfigurable patch in CST Microwave Studio. This antenna operates in two modes (bit-0 and bit-1), which are controlled by the two PIN diodes. In the scenario where PIN diode 1 is on and PIN diode 2 is off, the patch operates in bit-0 mode. Conversely, when PIN diode 1 is off and PIN diode 2 is on, the patch functions in bit-1 mode. The reflection coefficients of the two operating modes are shown in Fig. \ref{sim_patch}(a). As can be seen, the antenna operates at 10 GHz with a VSWR 2:1 impedance bandwidth of 360 MHz (3.6\%). The amplitude and phase of the radiation patterns of both modes in H-plane are shown in Fig. \ref{sim_patch}(b) and \ref{sim_patch}(c) in the $x-z$ plane at 10 GHz, respectively. Since the SLTM array introduced later is a linear array extending along the $x$-axis, SLTM will occur in the $x-z$ plane. Consequently, we only present the radiation patterns in the $x-z$ plane, for brevity. Leveraging the SPST characteristics of the PIN diodes, switching the diodes between ON and OFF states leads to an alternate switching in the direction of the surface current between forward and reverse bias modes of operation. However, the current's intensity remains constant, as visualized in Fig. \ref{current_flow}. This results in a radiation pattern exhibiting a $0^{\circ}$/$180^{\circ}$ phase shift without any change in amplitude. Thus, this patch can be used as an element of the SLTM array shown in Fig. \ref{SLTM_matlab} in which the excitation coefficients of 1 and -1 for the elements can be realized by controlling the state of the PIN diodes.

%
%
\begin{figure*}[h]
\centering
\includegraphics[width=\textwidth]{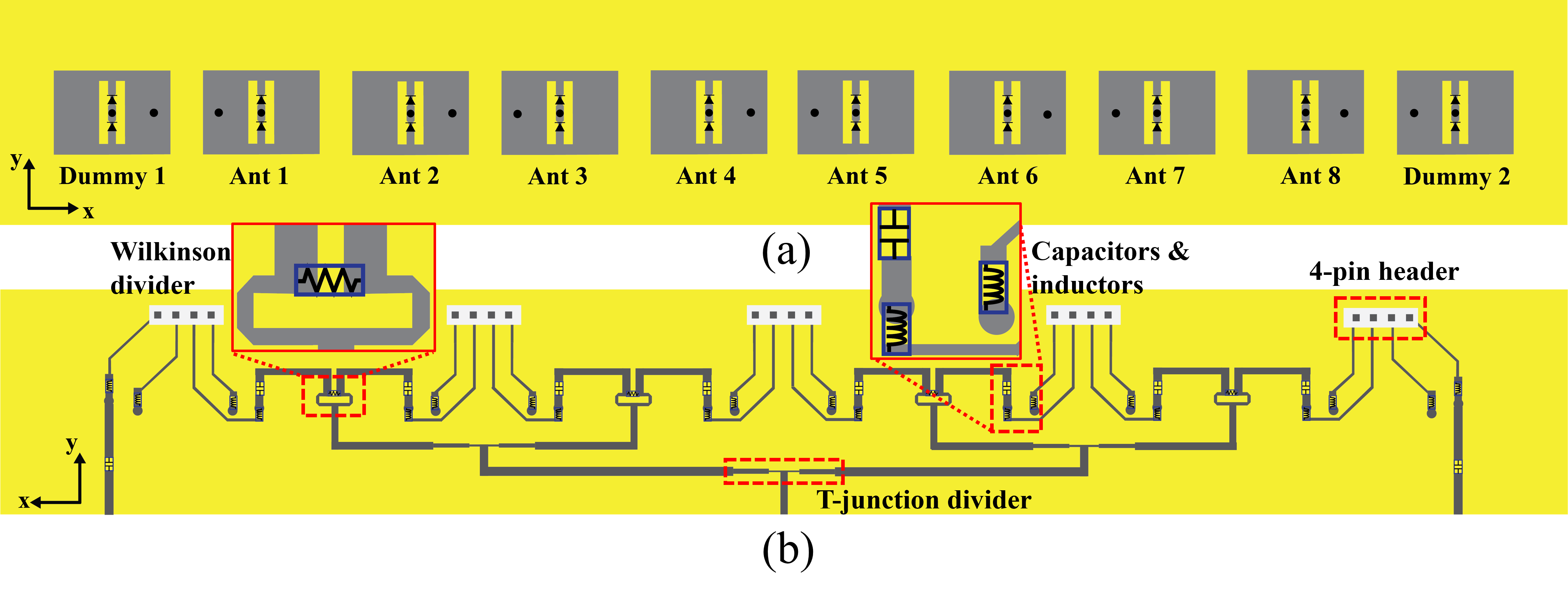}
\caption{ Topology of the SLTM array. (a) Top view. (b) Bottom view.}
        \label{array_model}
\end{figure*}


\subsection{SLTM Array Design}
To achieve the SLTM waveform developed in Section \ref{section2}, we designed an 8-element linear SLTM array using the 1-bit slotted path antenna elements discussed earlier. The topology of the proposed array is shown in Fig. \ref{array_model}, including both top and bottom views. This array consists of eight radiating patch elements (Ant 1-8), two dummy elements (Dummy1-2), and a feed network. The feed network consists of T-junctions and Wilkinson power dividers and is used to feed all antennas with the same amplitude and phase. The dummy elements positioned at the edges of the linear array are solely present to establish a periodic boundary along the $x$-axis and they play no role in beamforming. Without these dummy elements, the radiation patterns of Ant 1 and 8 would be different from those of the other radiating patches due to varying boundary conditions. These two dummy elements are loaded with 50 $\Omega$ loads and have no contribution to beamforming. The spacing between the center of each element is set to 17 mm (0.56 $\lambda_0$ at 10 GHz), this value slightly exceeds half wavelength and was chosen to  minimize mutual coupling between antenna elements. Minimizing the mutual coupling results in achieving surface currents for each element in the array that is not significantly changed compared to those of the isolated element shown in Fig. \ref{current_flow}. As a result, the patch elements can produce a clear 1-bit, phase-reconfigurable radiation pattern in the array environment as needed. Five 4-pin headers are used to connect the DC bias network of the SLTM array to an external FPGA that provided the bias voltages for the PIN diodes and controls their operating states.

The performance of the proposed SLTM array was evaluated using full-wave simulations in CST Microwave Studio. All eight elements were excited with the same amplitude and phase via the corporate feed network. By sequentially switching the PIN diodes on one of the eight elements to bit-1 and setting all other elements (including the dummy elements) to bit-0, we generated eight different radiation patterns corresponding to the array's eight operational modes. Fig. \ref{array_pattern}(a) illustrates the simulated reflection coefficients of these eight modes. Figs. \ref{array_pattern}(b) and \ref{array_pattern}(c) show the simulated amplitude and phase of the radiation patterns at 10 GHz. Notably, as the SLTM array cycles through its eight operational modes, the radiation patterns in the main lobe direction do not change significantly but the amplitude and phase of the radiation patterns in the sidelobe directions change considerably.

\subsection{Array Fabrication and Measurement}

\begin{figure}[h]
\centering
\includegraphics[width=0.5\textwidth]{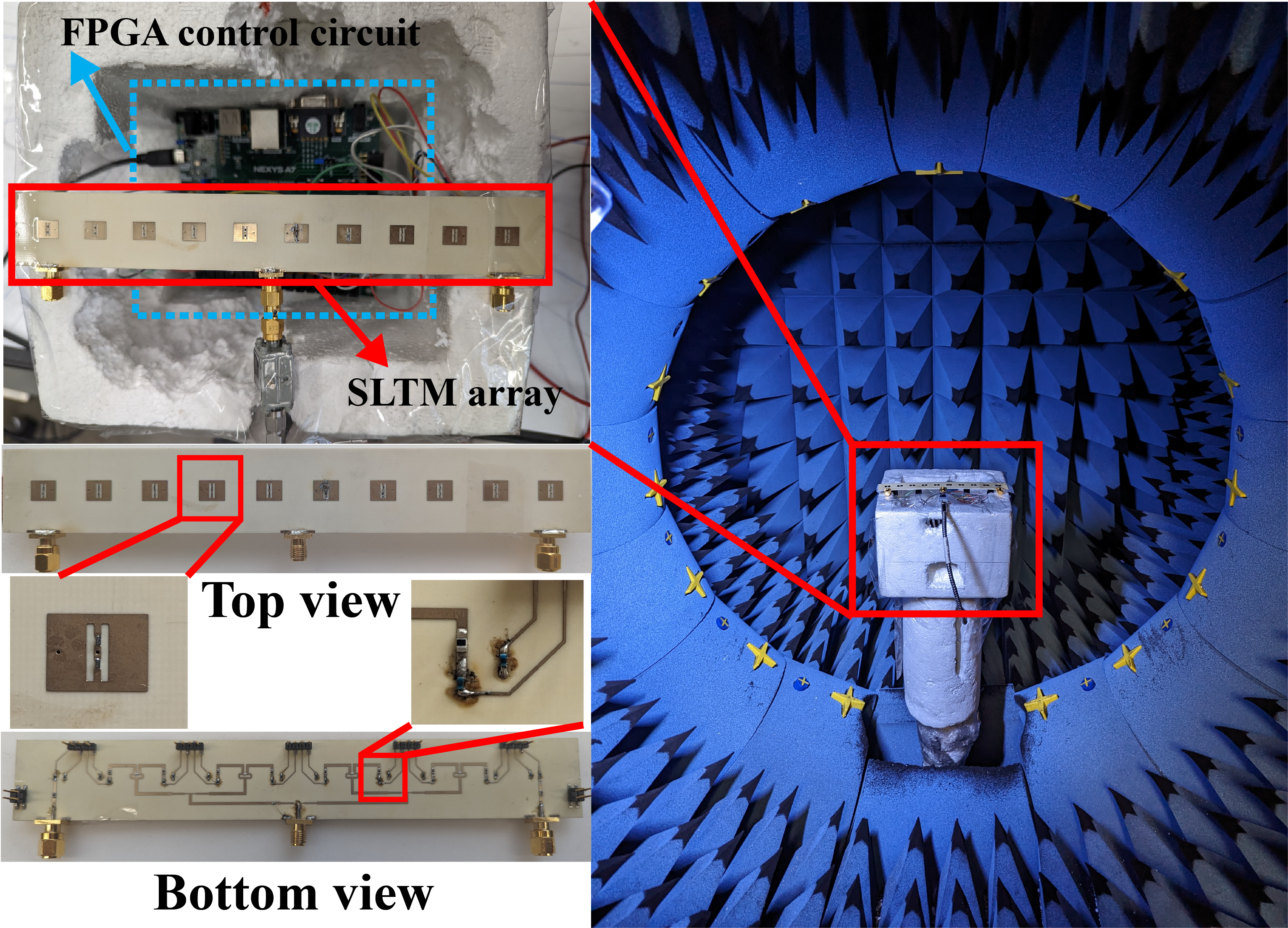}
\caption{{SLTM array fabricated prototype and radiation pattern measurment setup.}
\label{mea_prototype}}
\end{figure}

\begin{figure*}[h]
\centering
\includegraphics[width=\textwidth]{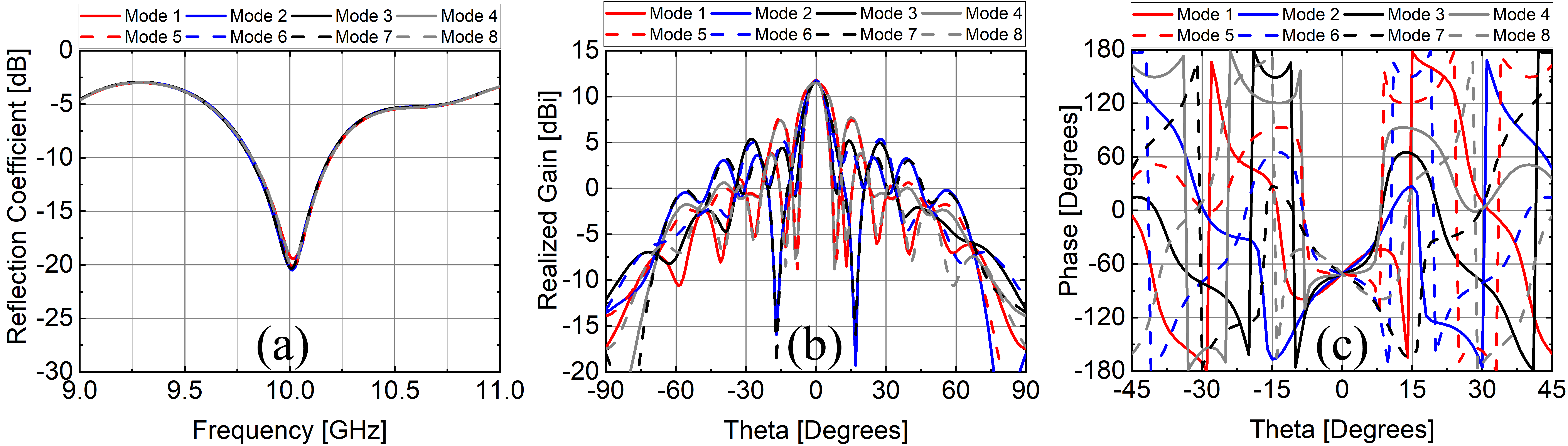}
\caption{(a) Simulated reflection coefficients of the SLTM array in different working modes. (b) Simulated amplitude and (c) phase of the radiation patterns of the SLTM array for different working modes simulated at 10 GHz.
\label{array_pattern}}
\end{figure*}

\begin{figure*}[h]
\centering
\includegraphics[width=\textwidth]{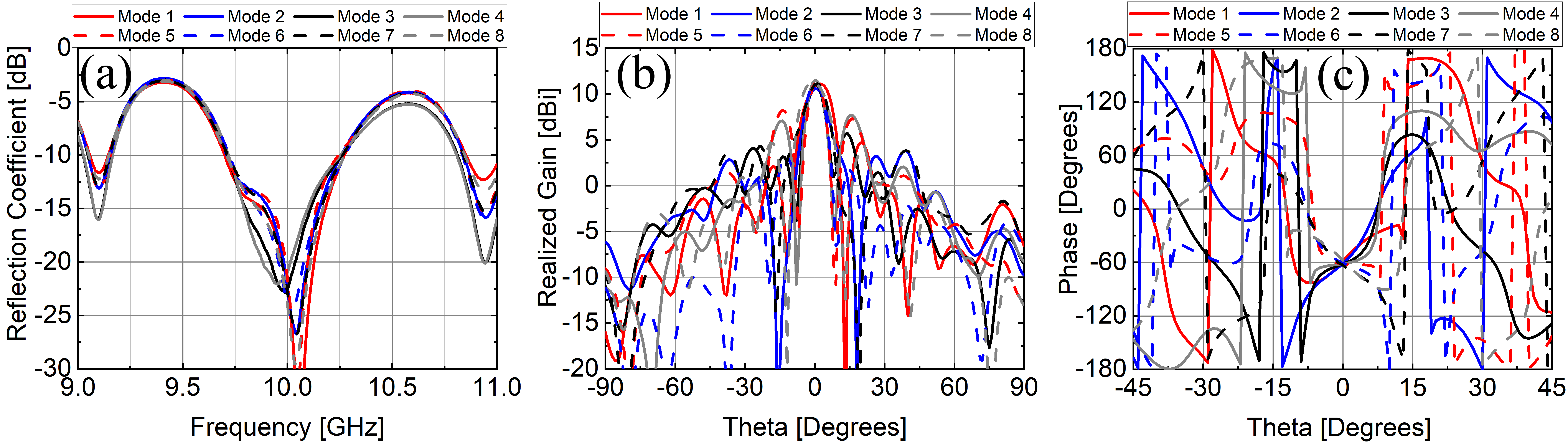}
\caption{(a) Measured reflection coefficients of the SLTM array in different working modes. (b) Measured amplitude and (c) phase of the radiation patterns of the SLTM array for different working modes simulated at 10 GHz.
\label{mea_result}}
\end{figure*}

To validate the principles of operation of the proposed SLTM array, the prototype discussed in the previous section was fabricated and experimentally characterized. Fig. \ref{mea_prototype} shows a photograph of the fabricated prototype and the test setup used for radiation pattern measurements. The reconfigurable SLTM array is controlled by an external FPGA (Nexys A7 from Xilinx) by dynamically changing the bias voltages on the PIN diodes of each element individually. The scattering parameters were measured using a power network analyzer, with two dummy ports each loaded with a 50 $\Omega$ load. The radiation patterns of the SLTM array were measured using the SATIMO StarLab system, with the measurement setup shown in Fig. \ref{mea_prototype}(b). Fig. \ref{mea_result}(a) presents the measured scattering parameters for the eight operational modes. In all operational modes, the array maintains a good impedance matching at 10 GHz, with a -10 dB $S_{11}$ bandwidth ranging from 9.75 to 10.2 GHz. The measurement and simulation results of Figs. \ref{mea_result}(a) and \ref{array_pattern}(a) agree well with each other. The measured amplitude and phase of radiation patterns are shown in Figs. \ref{mea_result}(b) and \ref{mea_result}(c). The measured radiation patterns are also consistent with the simulated ones and demonstrate that the radiation pattern of the array along the boresight direction does not change while the sidelobes change significantly for different modes of operation, as desired.

\section{Waveform setup and Numerical verification}
\label{section4}

\begin{figure}[h]
\centering
\includegraphics[width=0.45\textwidth]{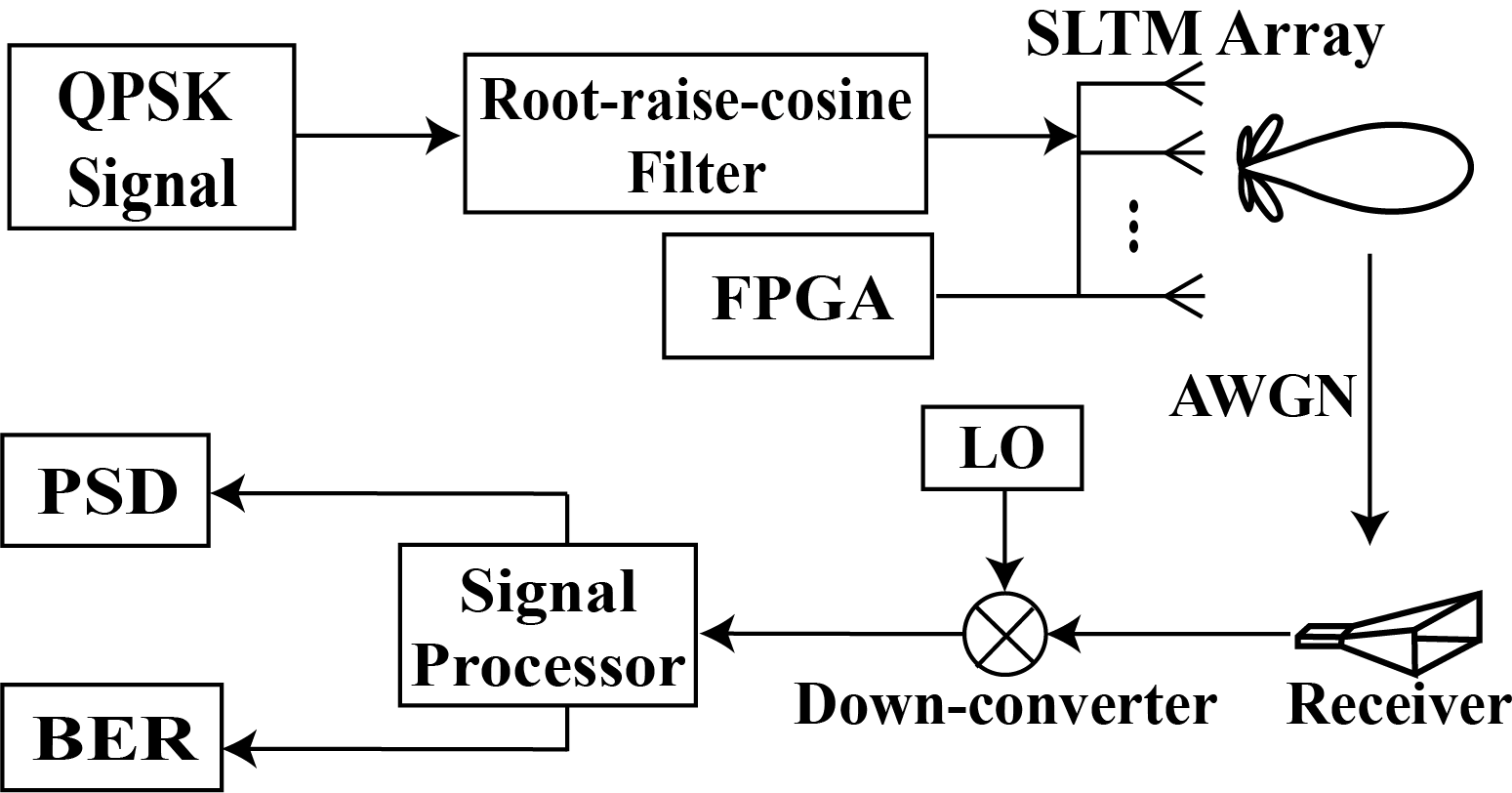}
\caption{Block diagram showing the setup used for simulating the PSD and BER of the SLTM array as part of a simple communications system. 
\label{simulation_diagram}}
\end{figure}

In this section, we examine the LPI/LPD and secure communications features offered by using the proposed SLTM array through numerical simulations in MATLAB. The measured amplitude and phase of the radiation pattern of the SLTM array are used in the numerical simulations. In these simulations, we assume that the SLTM array is used to transmit a QPSK-modulated signal and evaluate the PSD of the transmitted signal through the main lobe and sidelobes. We also evaluate the BER of the demodulated signal at different directions as a function of SNR to show that decoding the information that an eavesdropper may collect by picking up signals transmitted through the sidelobes of the SLTM array becomes more difficult. Then we study the SLTM in the receive mode where the desired signal is received from the main lobe direction, while the jamming signal is received from the sidelobe directions. We also use the BER of the received signal as a metric to evaluate the resilience to jamming. For a comprehensive study, we also evaluated the performance of a conventional eight-element array in the simulations alongside the proposed SLTM array as a baseline for comparison. Fig. \ref{simulation_diagram} shows the system-level block diagram of the setup used for performing these calculations when the SLTM works in the transmit mode.

In these simulations, we assume that the SLTM array is used to transmit or receive modulated signals centered at 10 GHz. We also assume that an eavesdropper attempts to detect the signals transmitted through the SLTM array's sidelobes and ascertain the presence of the transmitter, and after that deciphers the signal. Without loss of generality, we perform these simulations for a signal with QPSK modulation, although the same basic concepts can be applied to other modulation types as well. We assume that the QPSK signal has a bit length of $T_b$=0.5 µs and bandwidth of 4 MHz. On the receiver end the RF signal is down-converted to an intermediate frequency (IF) $f_{IF}$= 100 MHz which falls within the operational range of our signal processor. The operation frequency of the signal processor we used in the experiment is 0.07-6 GHz, which is lower than the carrier frequency of the transmitted signal. So in the experiments, the received RF signal at 10 GHz needs to be down-converted to $f_{IF}$ between 0.07-6 GHz, and we choose $f_{IF}$= 100 MHz. To be consistent, we also choose $f_{IF}$= 100 MHz in these simulations. The time modulation frequency,  $f_{SLTM}$, is set at 256 MHz. This ensures the signal is spread across a wide bandwidth, considering the performance constraints of the FPGA and the switching speeds of the PIN diodes employed in the SLTM array design.

\subsection{Design and Optimization of SLTM Sequence}

\begin{figure}[h]
\centering
\includegraphics[width=0.45\textwidth]{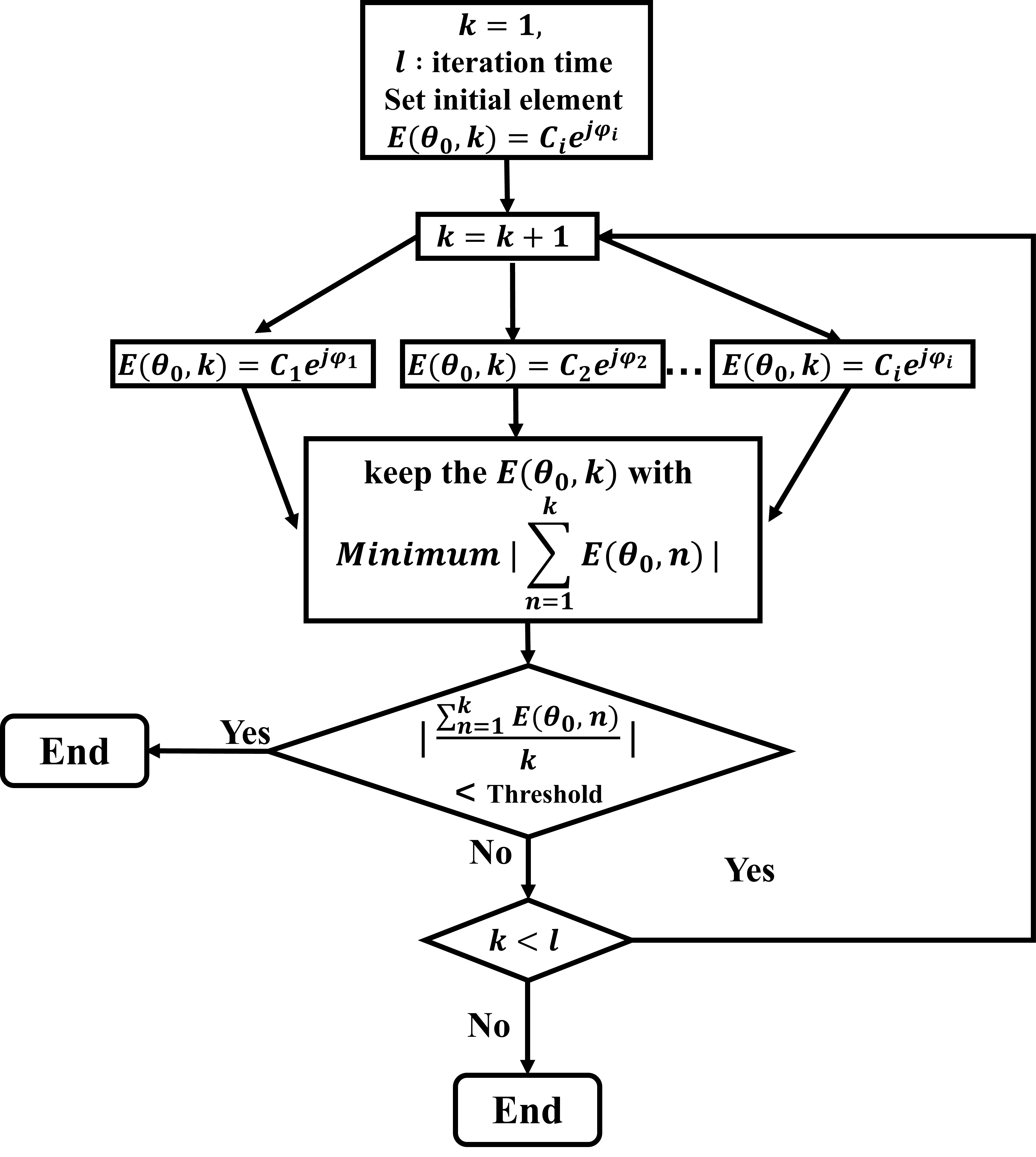}
\caption{Flow chart of the process used to optimize the SLTM time-modulated sequence.
\label{optimization_chart}}
\end{figure}

In Section \ref{section2}, we analyzed the PSD of the SLTM sequence, denoted as $P(f)$, and determined how to design the SLTM sequence $E(\theta_0,t)$ to achieve LPI/LPD secure communication and jamming resilience. Fig. \ref{optimization_chart} presents the flow chart of the TM sequence optimization process. Initially, we choose the starting value for the first element of the TM sequence. This value is selected randomly from the eight values in (\ref{pmf1}). After establishing the initial value, we simultaneously assign the next element with all eight values in parallel and find the one that minimizes $|\sum\limits_{n=1}^{k}E(\theta_0,n)|$. Next, we compare the expected value of $E(\theta_0,n)$ with a threshold value to determine if the optimization is complete or not. Ideally, the expected value should be 0 to completely eliminate the impact of the $\delta$ function. However, achieving an exact 0 is challenging using this optimization method. Thus, we set a threshold, and any expected value smaller than this threshold is considered acceptable. Our simulation and experimental results indicate that expected values smaller than $10^{-2}$ consistently provide effective suppression of the $\delta$ function and achieve convergence within a reasonable number of iterations. It is worth to note that this optimization procedure is only required to run once for a given antenna in the specific angle and can be used over and over.  In this optimized TM sequence, each of the eight values, $C_ie^{j\phi_i}$, has its corresponding proportion, $p_i$. In this paper, we do not use adaptive time modulation; instead, we pre-set the SLTM sequence in the FPGA, which repeats this sequence when the SLTM array is operational. This introduces periodicity into the SLTM sequence, but we mitigate this effect by making the preset SLTM sequence sufficiently long. In our simulations and experiments, we set the sequence length to $2^{13}$ and obtained satisfactory results, as presented later. However, the TM sequence obtained from optimization is usually not as long as the SLTM sequence we load into the FPGA. To address this, we extend the length of the TM sequence and then randomize all the elements in the extended sequence while keeping the proportion, $p_i$, of each value unchanged. Ultimately, this process yields the SLTM sequence for use in both simulation and experiments.

\subsection{Simulation of LPI/LPD Operation}

\begin{figure}[h]
\centering
\includegraphics[width=0.5\textwidth]{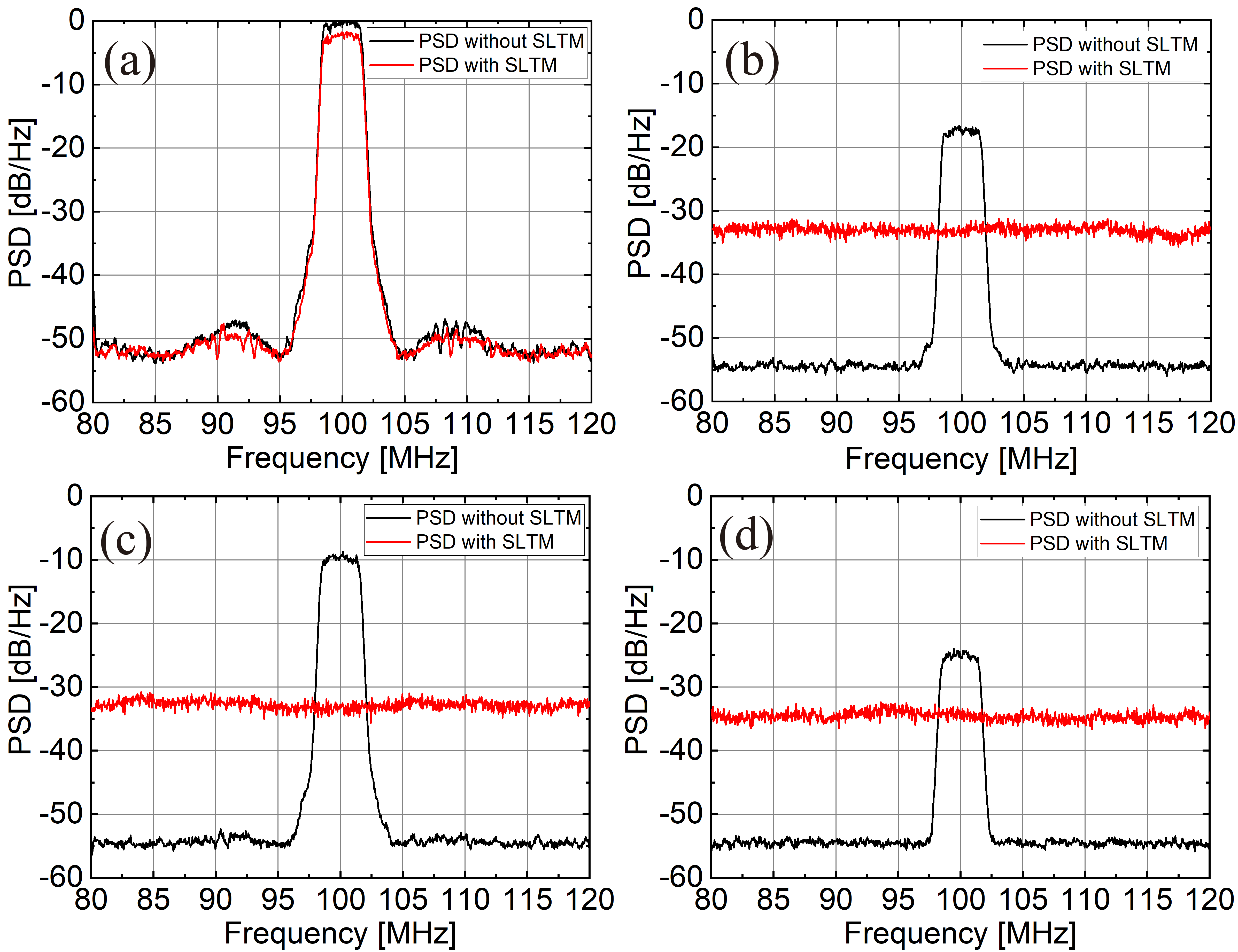}
\caption{Simulated PSDs of the conventional array and the SLTM array in both the main lobe ($0 ^{\circ}$) and the sidelobes ($10 ^{\circ}$, $20 ^{\circ}$ and $30 ^{\circ}$). (a) PSDs in the main lobe direction. PSDs along (b) $10 ^{\circ}$, (c) $20 ^{\circ}$, and (d) $30 ^{\circ}$.}
\label{PSD_sim}
\end{figure}

The PSDs for both the conventional and SLTM arrays are calculated at angles of $0^{\circ}$, $10^{\circ}$, $20^{\circ}$ and $30^{\circ}$. Here, $0 ^{\circ}$ represents the main lobe direction and the direction of the intended receiver, while the other angles denote undesired directions where potential eavesdroppers might be located. Fig. \ref{PSD_sim} illustrates the PSDs of the transmitted signal along these directions. All these PSD results are normalized to the PSD of the conventional array (i.e., without SLTM) along $0^{\circ}$. In the case of the conventional array, the signal is not scrambled in the sidelobe directions. Therefore, the shapes of the PSDs in the sidelobe directions are almost the same as that of the main lobe and only the transmitted power level decreases. Thus, an eavesdropper can easily detect the presence of the signal and identify the modulation type, perform signal classification and demodulation, and possibly localize the transmitter. For the SLTM array, the signal remains unscrambled in the main lobe, given that both the amplitude and phase of the radiation patterns remain constant during rapid switching across the eight operating modes. A slight power reduction of around 2 dB is observed in the PSD of the SLTM array when compared to that of the conventional array. This is attributed to the fact that only six elements participate in forming the main beam of the SLTM array in this example whereas in the conventional 8-element array, all eight elements are contributing to radiation towards the desired direction of radiation. Notice that as we increase the number of elements in a linear antenna array, this difference between the PSD of the SLTM array and that of the regular array in the desired direction of radiation becomes smaller. Fig. \ref{PSD_sim}(b)-\ref{PSD_sim}(d) show that the signal from the SLTM array undergoes significant distortion in undesired directions. Specifically, its PSD is noise-like over a 40 MHz band. Over a wider bandwidth, the spectrum of the signal transmitted through the sidelobes will be similar to that of a spread-spectrum signal. Additionally, depending on the eavesdropper location relative to the SLTM transmitter, the peak PSD value of the signal received by the eavesdropper is reduced by 10-25 dB relative to that of a conventional array. This reduces the effective range over which an eavesdropper can detect the presence of a transmitted signal, since at longer distances the received power may be below the noise floor of the eavesdropper's receiver system.

\subsection{Simulation of Secure Communication}
While diminishing the PSD can help in evading detection by eavesdroppers, the signal might still be discernible if its received power surpasses a certain threshold. Once detected, an eavesdropper might attempt to decode the signal and extract the embedded information. However, the signal in the frequency domain is completely distorted as depicted in Fig. \ref{PSD_sim}. This suggests that the original time-domain waveform is scrambled as well, making the decoding process challenging. Eavesdroppers would struggle to demodulate such a scrambled signal and retrieve any meaningful information from it. To quantitatively evaluate the communication security performance of the proposed SLTM antenna array, we have analyzed the BER of the QPSK signal it generates. For a comprehensive understanding, this BER is compared against the BER of a QPSK signal produced by a conventional eight-element array. The BER for the QPSK signal can be estimated using:
\begin{equation}
BER =\frac{1}{2}\text{erfc}(\sqrt{SNR})
\end{equation}

\begin{figure}[h]
\centering
\includegraphics[width=0.45\textwidth]{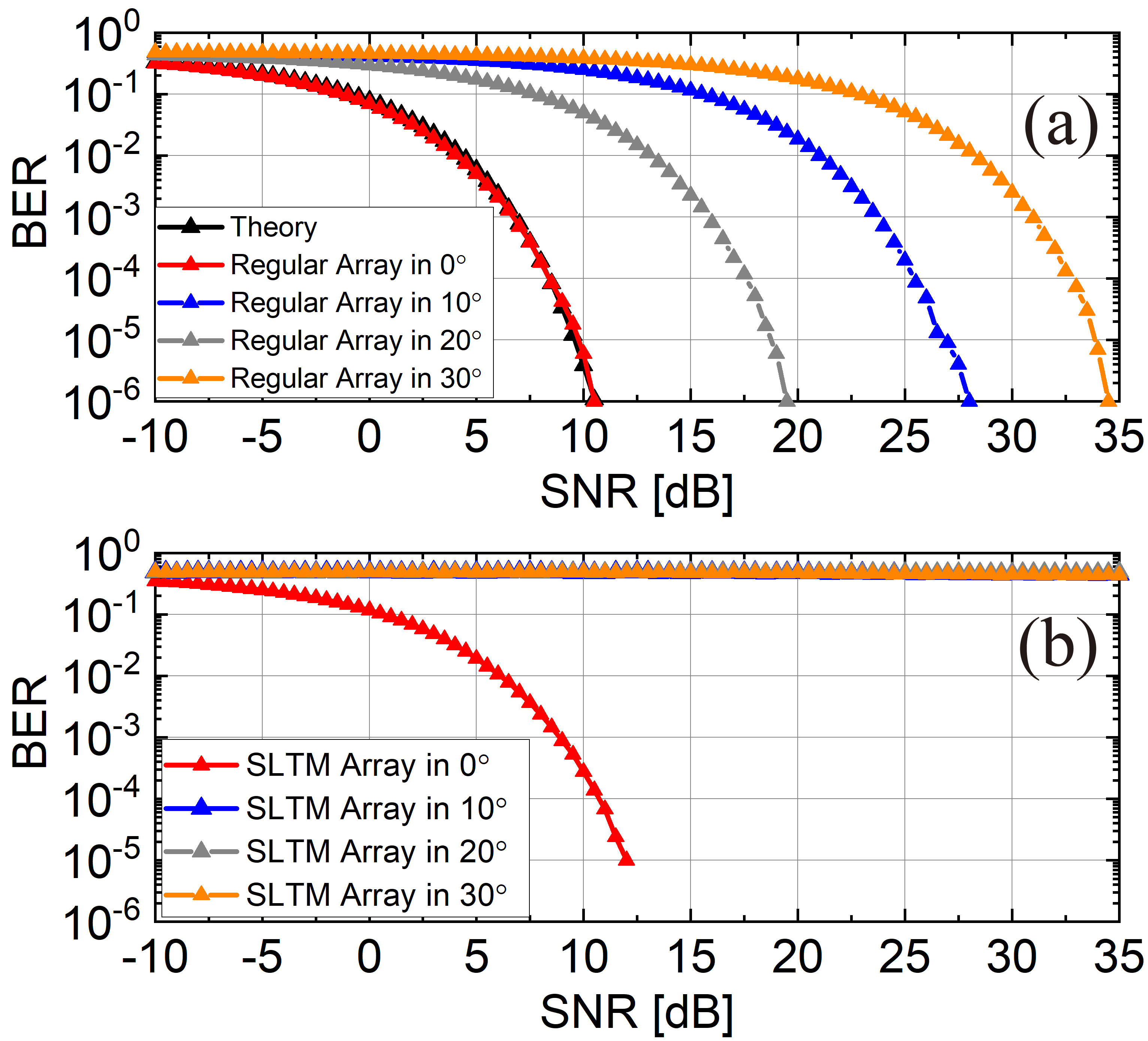}
\caption{Simulated bit error rate of (a) the conventional array and (b) the SLTM array, compared to the theoratical benchmark.
\label{BER_sim}}
\end{figure}

In communication systems, the Monte Carlo method is commonly used to simulate the BER. The theoretical BER of the signal with QPSK modulation is calculated as a reference. The SNR is calculated based on the transmitted power in the main lobe of the conventional array. As illustrated in Fig. \ref{BER_sim}, the BER from the conventional array aligns closely with this theoretical value. The BER from the SLTM array's main lobe has no change except a 2 dB shift in the SNR because of the PSD reduction of around 2 dB in the main lobe of the SLTM array compared to that of the conventional array. The BERs from the conventional array's sidelobes show similar properties as the BER from the SLTM array's main lobe but are only different in a shift in the SNR. These results show that waveforms in the main lobe of the SLTM array and all directions of the conventional array remain undistorted. On the other hand, the BERs from the SLTM array in the sidelobes approaching 0.5 means signals from the SLTM array's sidelobe undergo significant disruption due to the rapid pseudorandom time modulation. This perturbation results in the eavesdropper receiving a completely distorted information set.

\subsection{Simulation Verification of SLTM Array's Improved Resilience to Interference and Jamming}
\begin{figure}[h]
\centering
\includegraphics[width=0.47\textwidth]{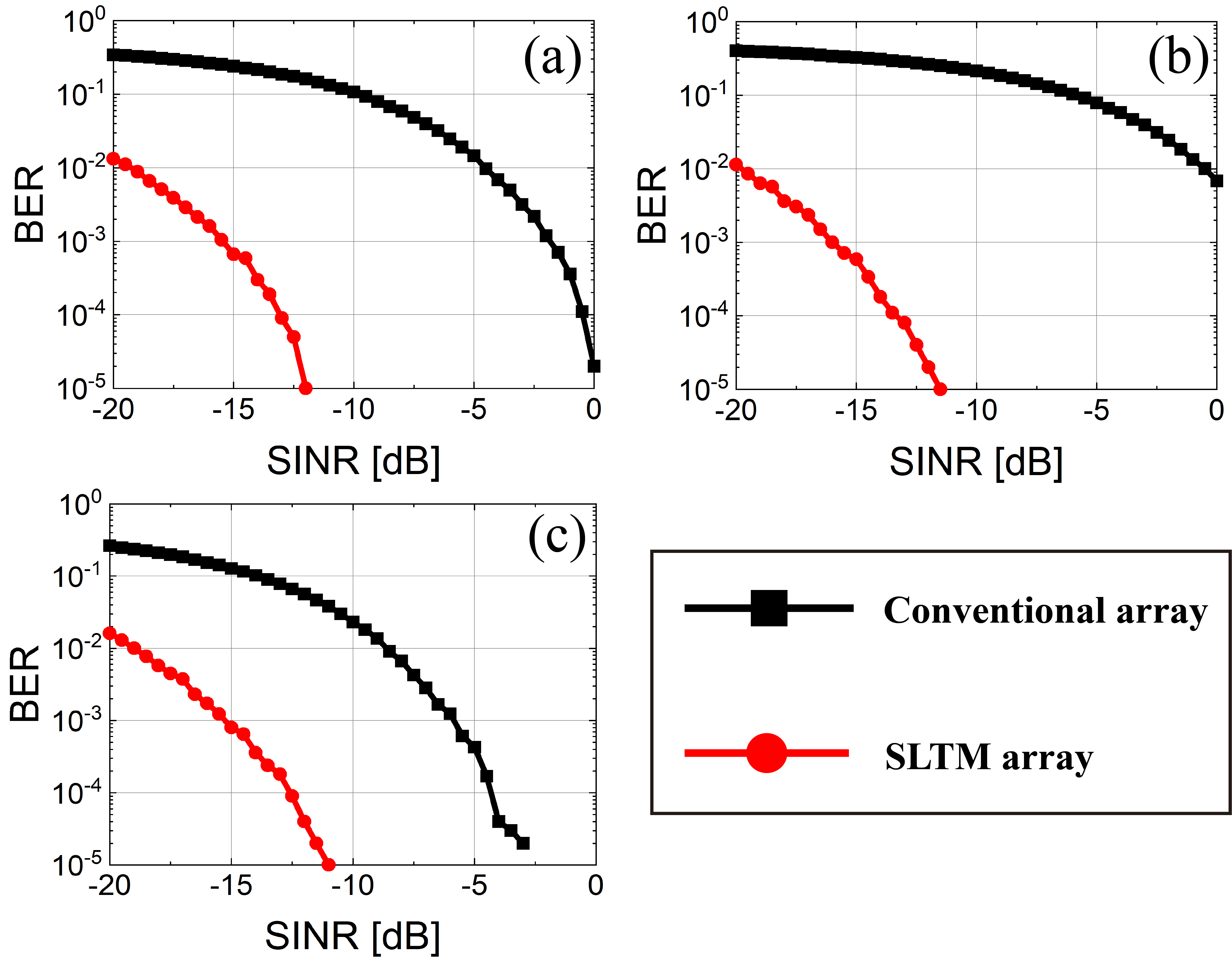}
\caption{Simulated bit error rate of the SLTM array in the receive mode with external jamming signal through the sidelobes. Bit error rates for the conventional and SLTM array are shown when the desired signal is received along the $0 ^{\circ}$ direction and the jammer is located along the (a) $10 ^{\circ}$, (b) $20 ^{\circ}$, and (c) $30 ^{\circ}$.
\label{simulation_BER_jamming}}
\end{figure}

In this subsection, we model the SLTM array in the receive mode and evaluate its ability to resist jamming and interference. In these simulations, the main lobe of the SLTM array points to the direction of the transmitter, and the jammer is assumed to be located at $10 ^{\circ}$, $20 ^{\circ}$ and $30 ^{\circ}$ sending additive white gaussian noise (AWGN) to the SLTM array. The signal-to-interference-plus-noise ratio (SINR) we define here is the power transmitted in the main lobe direction to the power of the interference transmitted from the jammer plus the white noise through the sidelobe direction of the conventional array. Fig. \ref{simulation_BER_jamming} shows the BER of the conventional array and the SLTM array jammed by external noise from undesired directions. For the same value of SINR, the BER of the SLTM array is lower than that of the conventional array, which means the SLTM array can reduce the impact of the interfering signal and improve the channel quality of the intended transmitter. 

The conducted simulations affirm the viability of the SLTM array in achieving LPI/LPD and secure communication, consistent with the theoretical analysis presented in Section \ref{section2}. It is noteworthy that this time modulation occurs at the antenna end, subsequent to signal encoding. Hence, regardless of the modulation, the PSD of the transmitted signal in undesired directions consistently aligns with (\ref{eqn8}). While our simulations used the QPSK modulation to demonstrate the principles of operation of an SLTM array, the approach remains versatile and can be extended to other modulations as well.

While a carrier frequency of 10 GHz was used for the preceding analyses, the SLTM approach presented is largely independent of the value of the carrier frequency since the carrier signal only shifts the spectrum of the modulated baseband signal to passband without impacting the spectral characterization. However, the values of $f_{SLTM}$ and $f_b$ must be carefully chosen to ensure desired performance. As per (\ref{eqn8}), if {$\E[B_m^{2}] \neq 0$} and $f_{SLTM}$ approaches $f_b$, the power of the SLTM sequence increases, leading to a higher PSD of the transmitted signal in undesired directions. Meanwhile, the bandwidth of the time-modulated sequence decreases and approaches that of the transmitted signal, thus the system no longer behaves in a spread-spectrum manner. This poses a challenge in distorting the characteristics of the PSD of the transmitted signal. To maintain LPI/LPD and enhance jamming resilience, $f_{SLTM}$ should exceed $f_b$, with greater values yielding better performance. Meanwhile, in our approach, we also choose the $f_{SLTM}$ greater than $f_b$ to enhance secure communications by using the spread spectrum modulation. However, even if $f_{SLTM}$ is close to or below $f_b$, the SLTM sequence can still scramble the information in the transmitted signal based on other operational principles \cite{Chen1_2022,z25} which are not investigated in this paper.

\section{Experimental Verification of LPI/LPD and Jamming Resilience Capabilities}
In this section, we conduct experiments to validate the LPI/LPD, secure communication, and jamming resilience of the proposed SLTM array. We utilize the SLTM array prototype discussed in  Section \ref{section3} in a communications system in both the receive and transmit modes. It is assumed that the broadside direction of the SLTM array aligns with the location of the intended receiver or transmitter, while the eavesdropper and jammer are situated off-axis, away from the main lobe.

\begin{figure}[h]
\centering
\subfigure[Subfigure 1 list of figures text][]{
\includegraphics[width=0.45\textwidth]{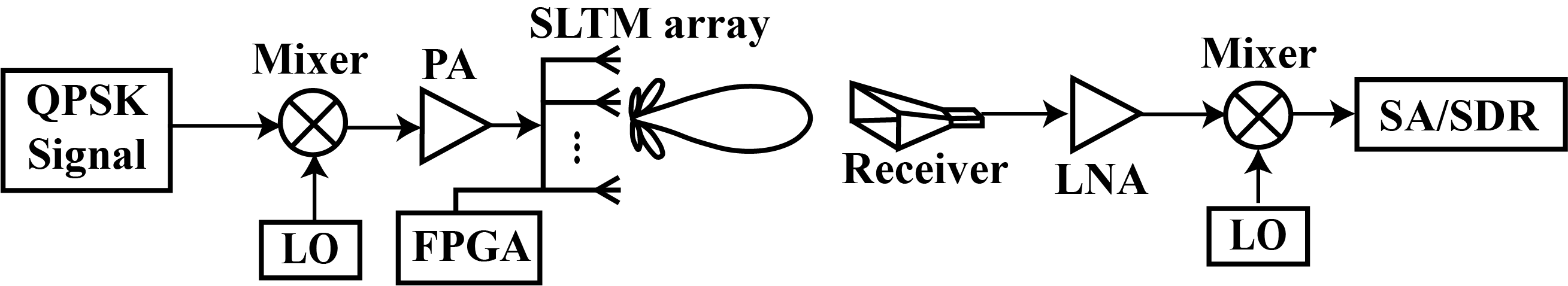}
\label{fig:subfig1}}
\subfigure[Subfigure 2 list of figures text][]{
\includegraphics[width=0.45\textwidth]{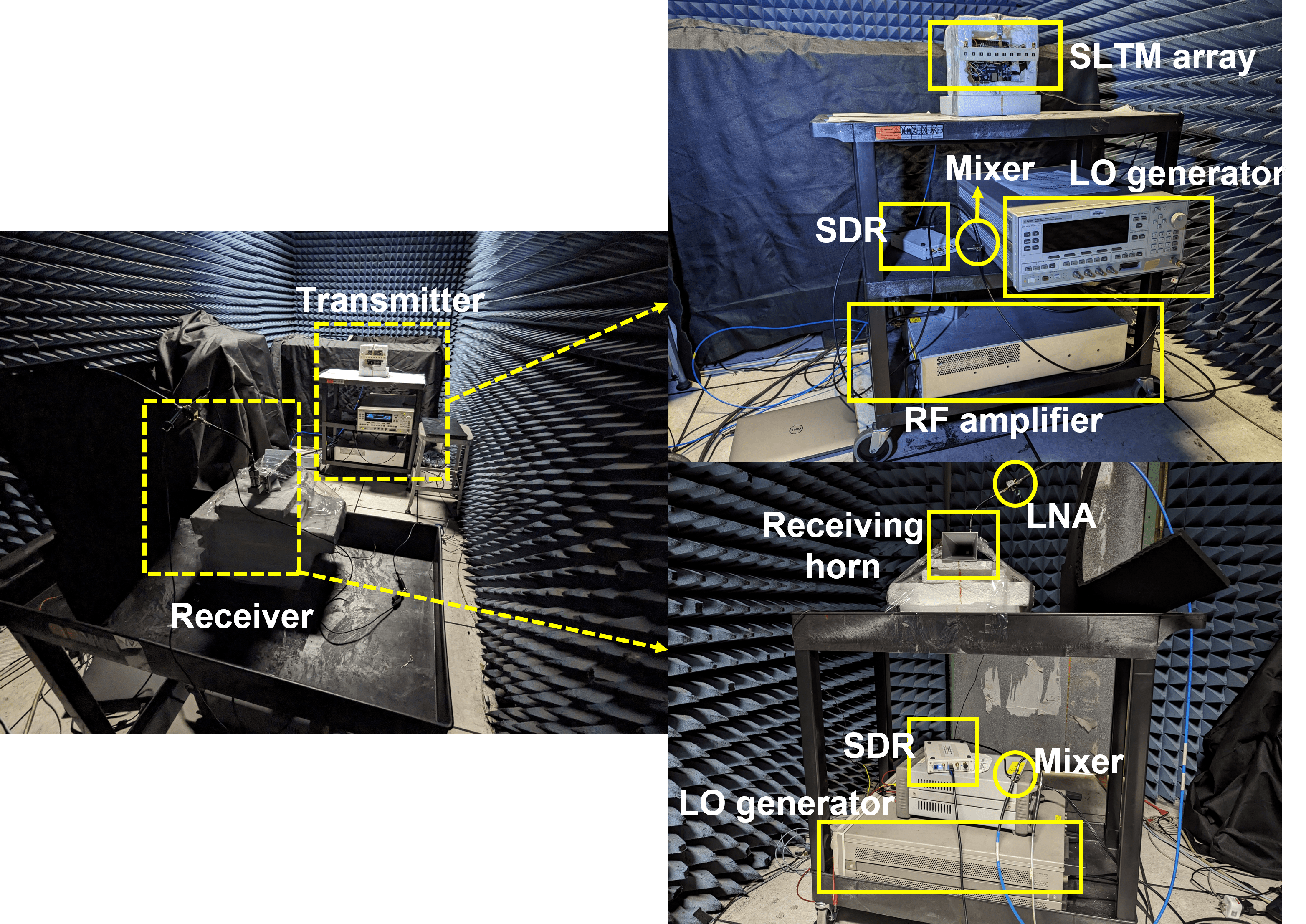}
\label{fig:subfig2}}
\caption{(a) Block diagram of the experimental setup employed for measuring the power spectral density and the bit error rate of a communication system employing SLTM array. (b) Photographs of the experimental setup.}
\label{experiment_setup}
\end{figure}

An experimental communication system is established to measure both PSDs and BERs, as illustrated in Fig. \ref{experiment_setup}(a). The specific experimental setup is further depicted in Fig. \ref{experiment_setup}(b). In this configuration, a software-defined radio (Ettus USRP-B210) is used to generate the modulated signal and upconvert it to the IF frequency. This signal is then upconverted to the carrier frequency (10 GHz) using a mixer, with the LO signal being generated from the signal generator (Keysight 83623B). Following this, the SLTM array, controlled by the FPGA, transmits the signal while time-modulating the sidelobes. At the receiver, the signal is downconverted to its IF frequency and subsequently analyzed using two instruments: an RSA306B spectrum analyzer and the Ettus USRP-B200 SDR, respectively. The spectrum analyzer is used to measure the PSDs of the received signal, while the SDR undertakes the demodulation of this signal, and calculates the BER.

%

\subsection{Power Spectral Density Measurements}
\begin{figure}[h]
\centering
\includegraphics[width=0.5\textwidth]{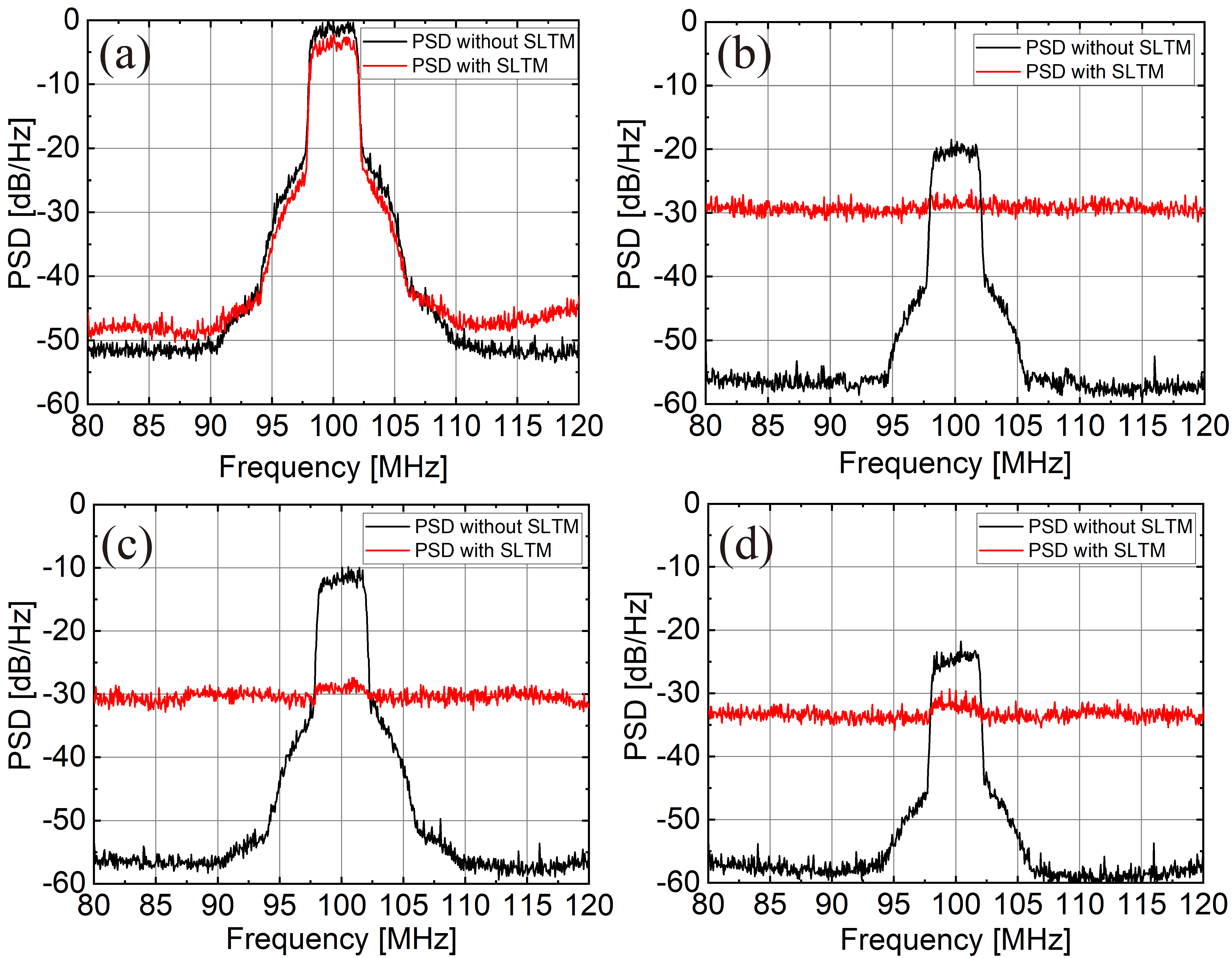}
\caption{Measured PSDs of the conventional array and the SLTM array in both the main lobe ($0 ^{\circ}$) and the sidelobes ($10 ^{\circ}$, $20 ^{\circ}$ and $30 ^{\circ}$). (a) PSDs in the main lobe direction. PSDs along (b) $10 ^{\circ}$, (c) $20 ^{\circ}$, and (d) $30 ^{\circ}$.}
\label{PSD_mea}
\end{figure}

Following the approach detailed in Section. \ref{section4}, we measured the PSDs of both the conventional and the SLTM arrays to validate the LPI/LPD performance of our proposed technique. Fig. \ref{PSD_mea} presents the PSDs of these arrays at angles of $0^{\circ}$, $10^{\circ}$, $20^{\circ}$ and $30^{\circ}$,  with the $0^{\circ}$ PSD of the conventional array serving as the reference. Observing the PSDs from the conventional array, the PSD in the sidelobe possesses reduced amplitude but retains a shape closely resembling the PSD in the main lobe. This suggests that, despite diminished amplitude, the QPSK signal waveform in unintended directions can still be easily detected by eavesdroppers and its properties classified. In contrast, the sidelobe signals from the SLTM array undergo significant scrambling, resulting in both reduced power and a completely different shape in the PSD. However, the main lobe signal from the SLTM array largely retains its characteristics, albeit with a slight amplitude reduction as explained before. These experimental results match those obtained from numerical modeling well.

\subsection{Experiment of Secure Communication}
\begin{figure}[h]
\centering
\includegraphics[width=0.48\textwidth]{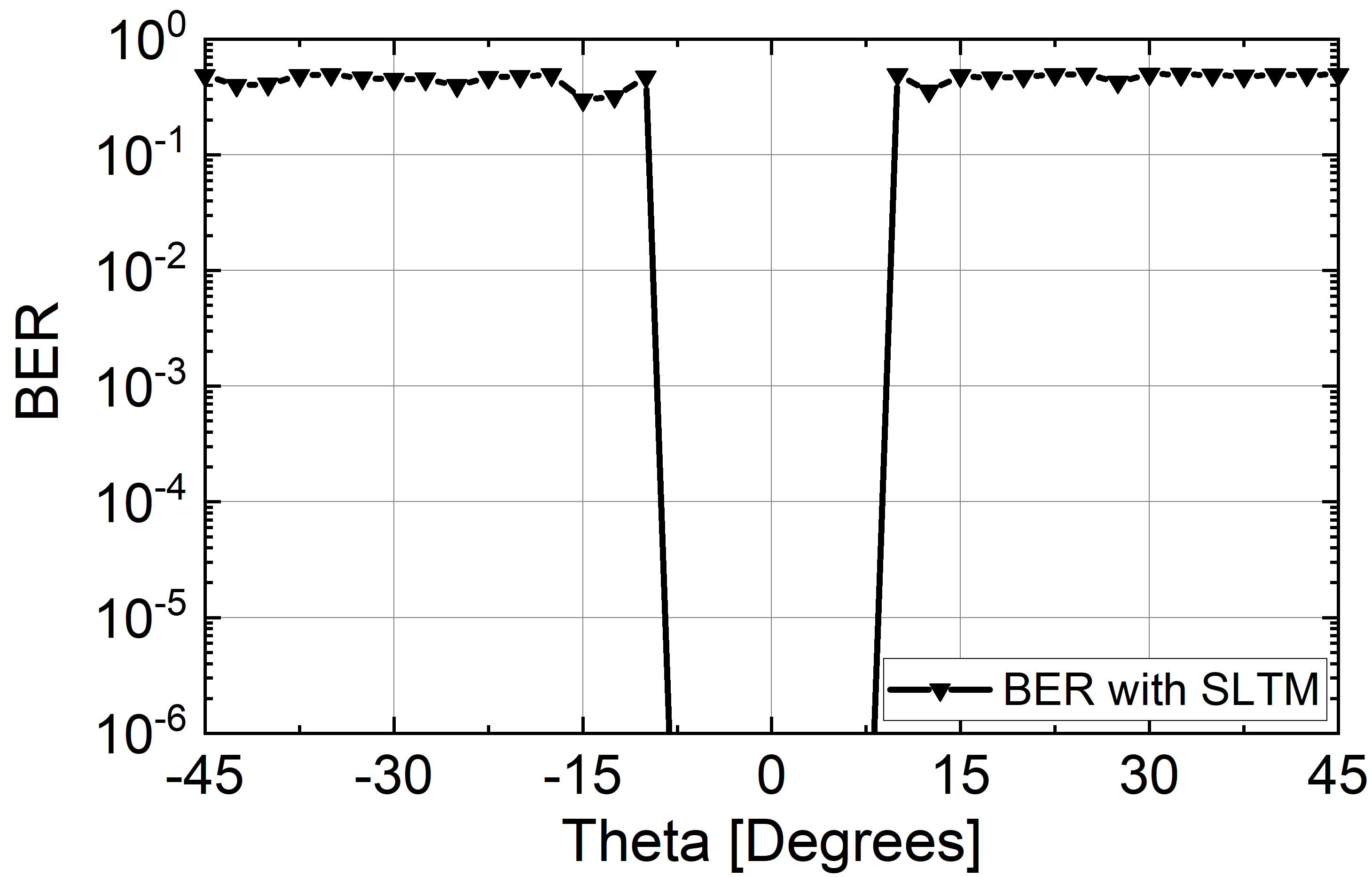}
\caption{Measured bit error rate at the receiver of a communications system employing the conventional array and the SLTM array as the transmitter. The receiver is moved from $-45^{\circ}$ to $45^{\circ}$. Since the SNR is up to 35 dB, the BERs we measured for the conventional array remain below $10^{-6}$ across the angle range from $-45^{\circ}$ to $45^{\circ}$.
\label{BER_mea}}
\end{figure}

To validate the difficulty of deciphering signals, which are transmitted through the sidelobes of the SLTM array, we measured the BERs of the transmitted signals. This evaluation spanned angles from $-45^{\circ}$ to $45^{\circ}$. In our experiments, an SNR of 35 dB was set in the main lobe direction of the conventional array. Then we fixed the power of the transmitted signal for the entire experiment of evaluating secure communication. This setting ensures that the measured BERs for the conventional array remain below $10^{-6}$ across the angle range from $-45^{\circ}$ to $45^{\circ}$. It is worth noting that when the BER goes below $10^{-6}$, an eavesdropper can conveniently demodulate the signal. Subsequently, the BERs for the SLTM array were assessed under identical conditions. The measurement results shown in Fig. \ref{BER_mea}, show that in the main lobe direction, the BERs remain below $10^{-6}$, indicating successful signal demodulation by the target receiver. Conversely, in the sidelobe direction, the BER values are all around 0.5. This suggests that the signals transmitted through the sidelobes behave similarly to noise, making it extremely challenging for eavesdroppers to demodulate and possibly decipher.

\subsection{Experimental Demonstration of SLTM Array's Improved Resilience to Interference and Jamming}
\begin{figure}[h]
\centering
\includegraphics[width=0.48\textwidth]{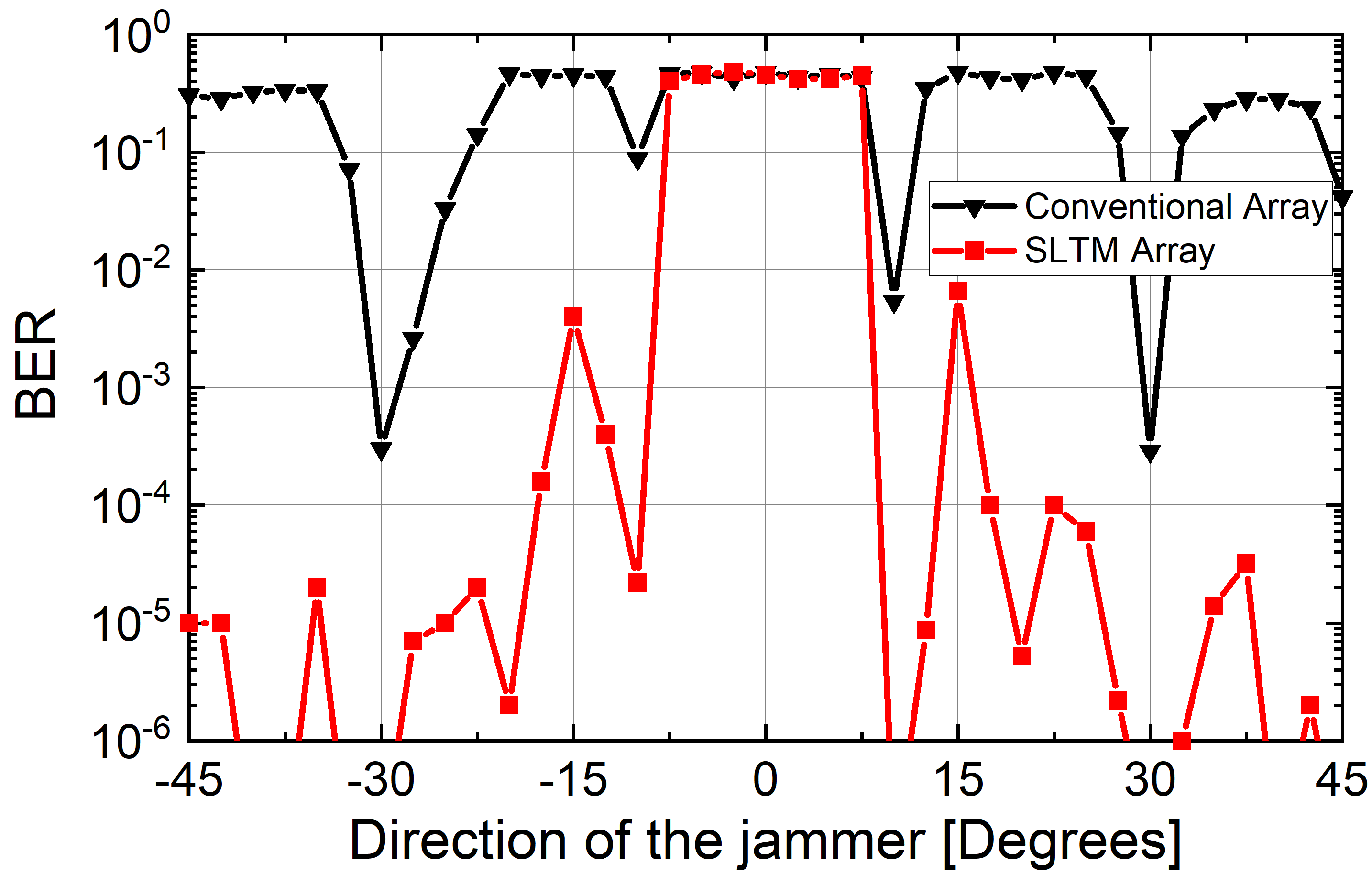}
\caption{Measured bit error rate of a communications system in which either the SLTM or the conventional array is used as the receiver and a jammer is present. The desired signal is transmitted from the boresight ($0^{\circ}$) and the direction of the jammer is varied from $-45^{\circ}$ to $45^{\circ}$. The signal transmitted by the jammer is 20 dB higher than the desired signal.
\label{BER_rx_mea}}
\end{figure}

In our LPI/LPD and secure communications experiments, the SLTM array served primarily as a transmitter. In this subsection, we shift our focus to evaluate the jamming resilience of the SLTM array when it is used in the receive mode. In the experimental setup, a transmitting horn was aligned with the main lobe direction of the SLTM array radiating a QPSK-modulated signal, which is the desired signal to be received by the SLTM array. A second transmitting antenna was used to transmit a strong interference signal towards the SLTM array and to emulate the presence of a jammer. The location of the jamming transmitter was systematically moved across a span of $-45^{\circ}$ to $45^{\circ}$. The radiated power levels of the desired and jamming transmitted signals were such that the desired transmitter provided a received SNR of 35 dB through the main lobe of the SLTM and that the jamming transmitter provided an SNR of 55 dB through the main lobe of the SLTM array. This provides a signal-to-interference ratio of -20 dB when both the desired transmitter and the jammer are aligned with the main lobe of the receiving SLTM array. The received signal through the SLTM array is then demodulated and the BER is calculated in the presence of this strong jamming signal for different locations of the jammer. The results, shown in Fig. \ref{BER_rx_mea}, reveal that as the jammer moves from $-45^{\circ}$ to $45^{\circ}$, the BERs for the conventional array approach a value of 0.5, except in specific directions where the conventional array's nulls are deep. However, the measured BER of the SLTM array outside of its main lobe is significantly improved compared to that of the conventional array even in the presence of such a strong interference. This shift can be attributed to the spreading of the power of the jamming signal over a broader bandwidth resulting in an improved signal to interference plus noise ratio. As elucidated in (\ref{eqn8}) the amplitude of PSD is influenced not only by the SLTM frequency ($f_{SLTM}$) but also by the inherent characteristics of the waveform itself ($B_m$). The SLTM array keeps the same $f_{SLTM}$ to reduce the power of the noise from all directions. However, the radiation pattern of the SLTM array varies in different directions and $B_m$ is dependent on the amplitude and phase of the radiation pattern. In this case, even though the SLTM waveform can curtail power in the frequency domain, the amplitude of this reduction varies in different directions. This causes the varied SINR and explains the variation in BER reductions across different directions.

\section{Conclusions}

{In this paper, LPI/LPD secure communications are achieved by developing the sidelobe time modulation method. By performing rapid sidelobe time modulation of an antenna array, the signals transmitted in the sidelobe directions are significantly distorted in a spread-spectrum-like manner, making them challenging to detect, identify, and decipher. However, the transmitted signal in the main lobe remains unchanged. In the receiving mode, through reciprocity, the power of a strong jamming signal received through sidelobes is reduced by spreading it over a wider bandwidth, reducing the efficacy of sidelobe jamming.} This provides a level of jamming and interference resilience to the systems using SLTM antenna arrays. We designed a linear array that provides time-dependent sidelobes and an unchanged main lobe to implement the SLTM method at X band. {The SLTM concept can be employed in existing communications systems by simply replacing their antennas with SLTM arrays without needing to make any other modifications to the system. This can bring LPI/LPD, communications security, and jamming resilience to many existing legacy communications systems that were not designed with these concerns in mind.} The operational principles of the SLTM method were demonstrated and verified through theory, simulations, and experiments. In the simulations and measurements, we used the PSD of the transmitted signal and the BER at the receiver to evaluate the proposed method. The measurement results agree well with the simulations. Although this paper primarily used the QPSK waveform for verification, the SLTM approach can also be applied to other digitally modulated waveforms. Each waveform undergoes time modulation at the antenna end, directionally scrambling the transmitted signals in undesired directions. In this work, we primarily focused on narrowband signals. The array factor of the SLTM array changes as the frequency changes. Therefore, the SLTM sequence optimized based on the array factors at one frequency may not be optimum for use at other frequencies. To extend the operation bandwidth of the SLTM method, the SLTM waveform needs to be optimized to work across the entire bandwidth of interest simultaneously. Studying this factor and the extension of the concept to more broadband signals are active areas of research that we hope to be able to report on in future works.

\appendices
\section{}

Since $u(t)$ is a stationary random signal, whose PSD can be expressed as the average power of the truncated function of $u(t)$ (Wiener–Khinchin theorem)
\begin{equation}
\centering
P_{ru}(f) = \lim_{T\to\infty} \frac{{\E[|U_T(f)|^{2}]}}{T}
\label{app4.1} 
\end{equation}
here $U_T(f)$ is the frequency domain response of the $u_T(t)$, $u_T(t)$ is the truncated function of $u(t)$, $T$ is the truncated time whose length is 
\begin{equation}
\centering
T=(M+1)T_{SLTM}
\label{app4.2} 
\end{equation}
here $M$ is a large integer, and substituting (\ref{app4.2}) into (\ref{app4.1}) we can get
\begin{equation}
\centering
P_{ru}(f) = \lim_{M\to\infty} \frac{{\E[|U_T(f)|^{2}]}}{(M+1)T_{SLTM}}
\label{app4.3} 
\end{equation}
To get $U_T(f)$, we calculate $u_T(t)$ from (\ref{ut}) and (\ref{ut2})
\begin{equation}
\centering
u_T(t)=\sum\limits_{m=0}^{M} u_m(t)= \sum\limits_{m=0}^{M} B_m  w(t-mT_{SLTM})
\label{app5.1} 
\end{equation}
then $U_T(f)$ is the Fourier transform of $u_T(t)$
\begin{equation}
\centering
\begin{aligned}
U_T(f) =\int_{0}^{\infty} u_T(t)e^{-2\pi ft} \, dt
\label{app5.2}
\end{aligned} 
\end{equation}
substituting (\ref{app5.1}) into (\ref{app5.2}) we can get
\begin{equation}
\centering
\begin{aligned}
U_T(f) &=\sum\limits_{m=0}^{M} B_m \int_{0}^{\infty} w(t-mT_{SLTM})e^{-2\pi ft} \, dt \\
          &= \sum\limits_{m=0}^{M} B_m e^{-j3\pi fmT_{SLTM}}\sinc(\pi fT_{SLTM})T_{SLTM}
\label{app5.3}
\end{aligned} 
\end{equation}
Thus
 \begin{equation}
\centering
\begin{aligned}
&|U_T(f)|^2 = U_T(f)U^{\ast}_T(f) = \sum\limits_{n=0}^{M}\sum\limits_{m=0}^{M}\\
&B_n B_m e^{-j3\pi f(m-n)T_{SLTM}}|\sinc(\pi fT_{SLTM})T_{SLTM}|^2
\label{app6.1}
\end{aligned} 
\end{equation}
The expected value of $|U_T(f)|^2$ is 
\begin{equation}
\begin{aligned}
\centering
&{\E[|U_T(f)|^2]}= \sum\limits_{n=0}^{M}\sum\limits_{m=0}^{M}\\
&{\E[B_n B_m]} e^{-j3\pi f(m-n)T_{SLTM}}|\sinc(\pi fT_{SLTM})T_{SLTM}|^2
\label{app6.2} 
\end{aligned} 
\end{equation}
when $m = n$
\begin{equation}
\resizebox{.5\hsize}{!}{$B_mB_n=B_m^{2}=\left\{
\begin{aligned}
 & b_1^2 , & \ p_1\\
 &b_2^2, & \ p_2\\
 & \qquad   \vdots & \\
 &b_i^2, & \ p_i
\end{aligned}
\right.$}
\label{app7.1}
\end{equation}
thus the expected value of $B_m^2$ is
\begin{equation}
\begin{aligned}
\centering
{\E[B_m^2]}&= \sum\limits_{i}^{}b_{i}^{2}p_i \\
                 &= \sum\limits_{i}^{}a_{i}^{2}p_i(1-P_i) - 2\sum\limits_{i}^{}\sum\limits_{q \neq i}^{}a_ia_qp_ip_q
\label{app7.2} 
\end{aligned} 
\end{equation}
while $m \neq n$, by using (\ref{bi}) , we obtained that {$\E[B_mB_n]$} is 0. Therefore, the expected value of $|U_T(f)|^2$ only exists when $m = n$. Thus (\ref{app6.2}) can be simplified as
\begin{equation}
\begin{aligned}
\centering
{\E[|U_T(f)|^2]}&=\sum\limits_{m=0}^{M}{\E[B_m^2]}|\sinc(\pi fT_{SLTM})T_{SLTM}|^2 \\
                    &=(M+1){\E[B_m^2]}|\sinc(\pi fT_{SLTM})T_{SLTM}|^2
\label{app7.3}
\end{aligned} 
\end{equation}
substuting (\ref{app7.3}) into (\ref{app4.3}) we can get the PSD of $u(t)$
\begin{equation}
\begin{aligned}
P_{ru}(f)&=\frac{(M+1){\E[B_m^2]}|\sinc(\pi fT_{SLTM})T_{SLTM}|^2}{(M+1)T_{SLTM}}\\
&=\left(\frac{{\E[B_{m}^{2}]}}{f_{SLTM}}\right)|\sinc(\pi fT_{SLTM})|^{2}
\end{aligned}
\end{equation}


%

\ifCLASSOPTIONcaptionsoff
  \newpage
\fi

\end{document}